\documentstyle[12pt]{article}
\def\theequation{\arabic{section}.\arabic{equation}}
\newcommand{\beq}{\begin{equation}}
\newcommand{\eeq}{\end{equation}}

\newcommand{\dd}{D\hspace{-.65em}/}

\newcommand{\bp}{\bar{\psi}}

\newcommand{\bm}{\bar{m}}
\newcommand{\bn}{\bar{n}}

\newcommand{\eq}[1]{Eq.~(\ref{#1})}
\newcommand{\rf}[1]{(\ref{#1})}
\def\bh{\bar{\phi}}

\def\h{\phi}
\def\p{\psi}
\def\c{\chi}
\def\bl{\bar{\lambda}}

\def\op{operator}

\def\zm{zero mode}

\def\cn{condition}
\def\LG{Landau-Ginzburg}
\def\sg2{\int_{\Sigma_2}\sqrt{g^{(2)}}}

\def\lsc{\partial_1\partial_{\bar{1}}+e^{-\rho} \partial_2\partial_{\bar{2}}}
\def\lvc{\partial_1\partial_{\bar{1}}+\partial_{\bar{2}} e^{-\rho}\partial_2}
\def\sc{e^{-\rho} \partial_2 \partial_{\bar{2}}}
\def\vc{\partial_{\bar{2}} e^{-\rho} \partial_2}

\begin{document}

\begin{titlepage}
\begin{flushright}
NBI-HE-94-34 \\
hep-th/y9407109\\
July 1994\\
\end{flushright}
\vspace{2.5cm}
\begin{center}
{\LARGE {\bf Infinite Conformal Algebras \\
in Supersymmetric Theories on Four Manifolds}}\\
\vspace{1cm}
\end{center}
\begin{center}
{\large Andrei\ Johansen}\footnote{E-mail: ajohansen@nbivax.nbi.dk \ / \
johansen@lnpi.spb.su \ }
\\ \mbox{} \\
{\it The Niels Bohr Institute,}\\
{\it Blegdamsvej 17, 2100 Copenhagen, Denmark}\\ \vskip .2 cm
and \\ \vskip .2 cm
{\it The St.Petersburg Nuclear Physics Institute,}\\
{\it Gatchina, 188350 St.Petersburg, Russian Federation}\\
\end{center}

\begin{abstract}
We study a supersymmetric theory twisted on a K\"ahler four manifold
$M=\Sigma_1 \times \Sigma_2 ,$ where $\Sigma_{1,2}$ are 2D Riemann
surfaces.
We demonstrate that it possesses
a "left-moving" conformal stress tensor on $\Sigma_1$
($\Sigma_2$) in a BRST cohomology, which generates the Virasoro algebra
with the conventional commutation relations.
The central charge of the Virasoro algebra has a purely geometric
origin and is proportional to the Euler characteristic $\c$ of the $\Sigma_2$
($\Sigma_1$) surface.
It is shown that this construction can be extended to include a
realization of a Kac-Moody algebra in BRST cohomology with a level
proportional to the Euler characteristic $\c .$
This structure is shown to be invariant under renormalization group.
A representation of the algebra $W_{1+\infty}$
in terms of a free chiral supermultiplet is also given.
We discuss the role of instantons and a possible relation between
the dynamics of 4D Yang-Mills theories and those of 2D sigma models.

\end{abstract}
\end{titlepage}
\newpage

\section{Introduction}
\setcounter{equation}{0}
Two dimensional
$N=2$ superconformal theories have been intensively studied in
recent years because of their relevance to superstrings, topological
theories, and integrable systems.
The large class of $N=2$ superconformal theories can be realized as the
infrared fixed points of the Landau-Ginzburg models
\cite{Ma,Va,LVW,howe,cec1,cec2}.
The difficulty of such a description is that the corresponding \LG \
models do not possess a conformal symmetry.
The proposal of an effective field theoretical description of $N=2$
superconformal theories in terms of \LG \ models has recently
received much further support.
Witten \cite{WiLG} has shown that under the operation of a half-twist the
\LG \ model turns out to be left-moving and conformally invariant.
This superconformal algebra is realized on the classes of cohomology
for one of the supergenerators of the $N=2$ SUSY algebra and turns out
to be invariant under the renormalization group.
This enables us to extract the essential information about the $N=2$
superconformal algebra realized at the infrared fixed point of the \LG \
model.
In particular Witten computed the corresponding elliptic genus.
Witten also extended his analysis to the case of $(0,2)$ SUSY models
coupled to abelian gauge multiplets \cite{SiWi}.
In Ref.\cite{mohri} this construction has been extended to more
complicated 2D $N=2$ models where $W_N$ algebras are realized in
the cohomology for one of the supergenerators.

The idea of identifying different theories in terms of appropriate
cohomologies was also used in \cite{MuVa} where
an equivalence of a special superconformal coset (with
$\hat{c} =3$) and $c=1$ matter coupled to two dimensional gravity
was demonstrated.
Moreover this coset was shown to be connected with a twisted version of
an euclidean two dimensional black hole in which
the ghost and matter systems are mixed.
This approach has also been used to identify bosonic strings and $N=1$
superstrings with particular classes of vacua for $N=2$ superstrings
\cite{BeVa} (for a further development see also \cite{BOP} and
the references therein).

A natural question is if it is possible to extend similar
constructions to the case of $D=4$ theories.
First, the existence of cohomological structures which are invariant under
the renormalization group can give some important information about
the dynamics of the theory.
Second, it has been observed \cite{WiLS}
that there is a relation between two-dimensional integrable
systems and four-dimensional self-dual Yang-Mills equations.
The mathematical conjecture \cite{ward1} (see also
\cite{ward2,mason,bak,abl,park,schiff,guil,castro} and the references
therein) that all possible bosonic
integrable systems in lower dimensions originate in the Yang-Mills
equations for self-dual connections (in $D=(2,2)$) provides an
additional motivation to look for such a correspondence.
It has been also demonstrated that
the supersymmetric integrable systems can be extracted from
supersymmetric self-dual Yang-Mills theory (see \cite{nishino} and
references therein).
An important element of
such an extraction is the dimensional reduction of four dimensional
self-dual Yang-Mills equations to two dimensions.

The important role of self-dual Yang-Mills connections has been supported
by recent results in $N=2$ superstring theory.
It is found \cite{OV,NiGa,siegel1} that the
consistent space-time background is $N=4$ self-dual supersymmetric
Yang-Mills theory for open $N=2$ superstrings, or $N=8$ self-dual
supergravity for $N=2$ closed (heterotic) superstrings \cite{siegel2}.
This correspondence
could reveal some hidden symmetries of string theory and help to
describe the structure of the string vacua from a unique symmetry
principle.

The importance of a relation between self-dual Yang-Mills connections
and two dimensional integrable systems for superstring theory
provides a motivation to look for such a
correspondence at the quantum level.

A possible tool for such a study is given by the topological theories.
An important example of such a theory is given by the topological
Yang-Mills theory \cite{witten} which is a twisted version of
the $N=2$ supersymmetric Yang-Mills theory.
The space of physical \op s in the topological Yang-Mills theory is
defined as cohomology classes of an appropriate BRST \op .
The action of the theory turns out to be BRST exact and, hence,
the physical correlators
are independent of the external metric and the gauge coupling constant.
Therefore they
can be calculated semiclassically around instanton configurations
which turn out to be an essential ingredient of the theory because the
physical correlators can be represented as integrals over the instanton
moduli space.
These correlators can depend on moduli of
a differential structure on the curved manifold and
give a realization of the Donaldson map \cite{don} that relates
a smooth structure of a 4-dimensional manifold to the topology of the
instanton moduli space in terms of physical correlators.
The physical correlators turn out to be Donaldson's invariants which
characterize smooth structures on 4D-manifolds
\footnote{The topological Yang-Mills
theory is also intriguing from the physical point of view.
In particular in the heterotic string theory the scattering
amplitudes of spacetime axions at zero momenta are found to be proportional to
the Donaldson invariants \cite{harvey}
in the form as they are presented in Witten's theory \cite{witten}.}.

Notice however that the topological Yang-Mills theory \cite{witten}
has a finite number of physical degrees of
freedom as a result of a reduction of the space of physical states of
the $N=2$ supersymmetric theory by using the BRST \op .
In turn
a four dimensional analogue of 2D half-twisted theories can be
an intermediate situation between the usual dynamical 4D theories and
4D topological theories and provide us with a sort of a quantum
dimensional reduction.

Recently \cite{ajoh} it has been shown that $N=1$ $D=4$ SUSY gauge
theories with an appropriate representation of matter can be twisted
on a K\"ahler manifold.
These twisted models have an appropriate BRST charge,
which is one of the N=1 supergenerators.
The BRST charge does not depend on the external metric.
If the theory does not contain any superpotential for the matter
supermultiplet then the action is BRST exact.
Otherwise it is BRST closed.
The physical \op s are
defined as the classes of cohomology of the BRST \op .

Such a twisting is a four dimensional analogue of a half-twisting
of 2D theories \cite{WiLG}.
The topological Yang-Mills theory \cite{witten} is actually a
particular case of such a twisted model with the fields of matter in
the adjoint representation of the gauge group.
However in the topological Yang-Mills theory
\cite{witten} and in its twisted $N=1$ supersymmetric version
one considers the cohomologies of the different BRST \op s
\footnote{Recently \cite{witten4}
Witten calculated the Donaldson invariants by using
the formulation of the topological Yang-Mills theory \cite{witten}
in terms of the twisted N=1 supersymmetry.}.
An extension of the construction of ref.\cite{ajoh} to supersymmetric
theories without gauge interactions is straightforward (we illustrate it
in the present paper).
We will henceforth generically name the twisted $N=1$ supersymmetric
theories heterotic topological theories.

The space of physical \op s in the heterotic theory is in general
more complicated than in the topological Yang-Mills theory.
This space contains a ground ring of the local operators which are BRST
invariant off-shell.
This is a purely topological part of the model and this ring coincides
with the ring of local physical \op s in the Witten's topological Yang-Mills
theory provided that the matter is in the adjoint representation of the gauge
group.
However there is also an infinite set of \op s in the cohomology
of the BRST \op \ which can depend on the external metric and commute with
the BRST \op \ only up to the equations of motion.
This structure of the space of physical \op s is similar to that of
2D half-twisted models \cite{WiLG,SiWi}.
In general the correlators with insertions
of such \op s depend on the external metric
due to a presence of the external metric in the inserted \op s.
They can also depend on holomorphic coordinates on the four dimensional
K\"ahler manifold \cite{ajoh} but do not depend on the anti-holomorphic
ones.
The latter fact implies that the twisting procedure provides us with a
dimensional reduction from four dimensional space into two dimensional
one.

In turn in
the heterotic topological gauge theories (without a superpotential)
the physical correlators allow
for a localization near the solutions of equations of motion in the path
integral due to the BRST exactness of the action
and, hence, are determined semiclassically by the fluctuations near
instantons similar to the topological Yang-Mills theory.
Therefore we may expect that the heterotic topological gauge
theory can give a relation between self-dual connections and 2D
integrable systems at the quantum level.

In the present paper we focus to the theories on
a four manifold $M=\Sigma_1 \times \Sigma_2 ,$ where
$\Sigma_{1,2}$ are 2D Riemann surfaces.
We show that the twisted supersymmetric
theory on a four manifold $M=\Sigma_1 \times \Sigma_2$
possesses two left-moving conformal stress tensors
which generate two chiral Virasoro algebras (on $\Sigma_1$($\Sigma_2$))
in the cohomology of a BRST \op \ \cite{joh}.
The central charges of these algebras are shown to be invariant under the
renormalization group and can be calculated in the weak
coupling limit.
These central charges
turn out to be proportional to the Euler characteristic of a Riemann
surface $\Sigma_1 (\Sigma_2)$.
In addition a twisted theory can contain BRST closed holomorphic (in a BRST
cohomology) currents which generate a Kac-Moody algebra on
$\Sigma_1$($\Sigma_2$) with a level proportional to the Euler
characteristic of $\Sigma_2$($\Sigma_1$).
We also show that the $W_{1+\infty}$
\cite{wintro,center,bakas} algebra can be realized in terms of
a free chiral supermultiplet.

It is interesting that the central charge manifests itself as a
gravitational anomaly of the four dimensional twisted theory.
Actually we demonstrate that due to
this anomaly two Liouville theories (for the external metric)
are generated on the surfaces $\Sigma_{1,2} .$
It turns out that the multiloop corrections to the gravitational anomaly
vanish due to the BRST invariance of the theory.
Therefore we are tempted to look at the effective two dimensional
conformal theories from the point of view of two-dimensional gravity.

It turns out that in the case of matter in the
adjoint representation of the gauge group the central charge of the
Virasoro algebra $c=0.$
A tempting possibility is that one could
identify this formulation (twisted N=1 supersymmetry) of the
topological Yang-Mills theory with a topological
2D gravity coupled to topological matter in terms of a BRST
cohomology.

One can also hope that such a construction can give an important information
on the renormalization group (RG) in $D=4$ $N=1$ super QCD.
We discuss below the instanton effects in the heterotic
topological Yang-Mills theory and
demonstrate an appearance of a possible correspondence between
RG flows and instantons in 4D super QCD and in 2D sigma models.

The paper is organized as follows.
In section 2 we review the twisting procedure of supersymmetric theories
on K\"ahler manifolds and formulate the heterotic topological theory.
This construction is also extended to self-interacting chiral supermultiplet
without gauge interactions.
We define the physical \op s as classes of the BRST cohomology.
We discuss the anomalies in a decoupling of the external metric from the
physical correlators and the \cn s of the anomaly cancellation and the
renormalization properties of the gravitational anomaly.
In section 3 we demonstrate a realization of the $W_{1+\infty}$ algebra
in terms of a free chiral supermultiplet in the ghost number $G=0$ BRST
cohomology.
We also briefly discuss the BRST cohomology at $G\neq 0 .$
In section 4 we give a realization of the Virasoro algebra in terms of
a self-interacting chiral supermultiplet.
In section 5 we focus to twisted supersymmetric gauge theories.
We discuss the procedure of gauge fixing and the properties of physical
correlators.
We then study the BRST cohomology at the ghost number $G=0$ from the
point of view of $U(1)$ and Konishi anomalies.
We give a realization of Kac-Moody and Virasoro algebras.
We finally discuss the instanton effects and a possible relation of the
dynamics of the four dimensional Yang-Mills theory to that of two
dimensional sigma models.
We conclude by summarizing the results of the paper.
Some possible generalizations of these results are also discussed.

\section{Supersymmetric theory on a four dimensional
K\"ahler manifold}
\setcounter{equation}{0}

\subsection{Coupling to a reduced multiplet of supergravity}

Let us consider $N=1$ supermultiplet in a euclidean compact
four manifold $M .$
On a curved manifold all the supersymmetries are broken by an external
metric.
However as it is shown in \cite{ajoh}, on a K\"ahler manifold, if we
introduce an external vector field $V_{\mu}$ coupled to an appropriate
axial fermionic current constructed out of quantum fields
one of the four supergenerators is associated with
an unbroken fermionic symmetry
of the theory even for non-trivial external K\"ahler metric.
This vector field should depend on the external metric and is fixed by
the \cn \ that one of the supergenerators survives in the curved metric
(for a similar approach to the twisting of the $N=2$ supersymmetric
theories see \cite{ym,kr,galperin}).

In order to implement
twisting it is convenient to consider a supermultiplet
coupled to an external $N=1$ supergravity multiplet that contains
the veirbein field $e^a_{\mu} ,$ the gravitino (Weyl spinor)
field $\chi_{\mu} ,$ $\bar{\chi}_{\mu} ,$
$U(1)$ vector field $V_{\mu}$ and auxiliary fields.
We consider these supergravity fields as external ones.
The total system is invariant under simultaneous local
supertransformations both of matter and supergravity fields.
In a second step we must reduce the external supergravity multiplet in order
to get a supergravity background which is invariant under a global
supertransformation.
It turns out that such a reduction
is possible only on K\"ahler manifolds \cite{ajoh}.
We carry out the reduction to the veirbein field
$e^a_{\mu}$ and the vector field $V_{\mu} .$
By using the formulas of supertransformations of the ("new minimal")
supergravity \cite{west} we observe that this reduced supermultiplet
is invariant under a SUSY transformation if the variation of the
gravitino field vanishes
\beq
\delta \chi_{\mu} = (\partial_{\mu} - \frac{i}{4} \omega^{ab}_{\mu}
\sigma_{ab} + V_{\mu})\epsilon ,
\eeq
$$\delta \bar{\chi}_{\mu} = (\partial_{\mu} - \frac{i}{4} \omega^{ab}_{\mu}
\sigma_{ab} - V_{\mu})\bar{\epsilon} .$$
Here the letters $\mu ,$ $\nu ...$ stand for world indices while
$a,$ $b...$ correspond to the Lorentz indices in the tangent frame;
$\sigma^{ab} = [\sigma^a ,\sigma^b]/2i ,$ where $\sigma^a ,$ $a=1,2,3,4$
are the Pauli matrices.
The left- (right-)handed spinors $\epsilon$($\bar{\epsilon}$) are
parameters of the supertransformation.
The spin-connection $\omega^{ab}_{\lambda}$ reads
\beq
\omega^{ab}_{\lambda}= \frac{1}{2} [ e^{a\nu} e^{b\mu} e_{c\lambda}
(\partial_{\nu} e^c_{\mu} -\partial_{\mu} e^c_{\nu}) -
e^{a\nu}
(\partial_{\lambda} e^b_{\nu} -\partial_{\nu} e^b_{\lambda})-
e^{b\nu}
(\partial_{\nu} e^a_{\lambda} -\partial_{\lambda} e^a_{\nu})] .
\eeq
On the K\"ahler manifold the holonomy group is reduced to $U(2)$
(see, e.g. \cite{green}) and in the sector of right handed spinors
we can choose for definiteness
the matrix $i\omega^{ab}_{\mu} \sigma_{ab}/4$ to be
$\omega_{\mu} \sigma_3$ (with some $\omega_{\mu}$) for all $\mu$
in an open region of the 4-manifold.
We can then take $V_{\mu} =\pm \omega_{\mu} .$
The global supersymmetry that
survives in this background is generated by one of the right-handed $N=1$
supercharges and corresponds to a constant parameter $\bar{\epsilon} .$

{}From now on we shall consider the reduced supergravity multiplet with
the veirbein field and the vector $V_{\mu}$ obeying
\beq
(\frac{i}{4} \omega^{ab}_{\mu} \sigma_{ab} + V_{\mu})\bar{\epsilon} =0,
\label{defeps}
\eeq
for a non-vanishing constant spinor $\bar{\epsilon} \neq 0.$

Let us first consider the gauge supermultiplet coupled to the external
metric and the vector field $V_{\mu} .$
The $N=1$ gauge supermultiplet contains a gauge field $A_{\mu} ,$
a fermionic gaugino field $\lambda_{\alpha}$ and
$\bar{\lambda}_{\dot{\alpha}} ,$ and
an auxiliary scalar field $D$ which is necessary for SUSY algebra to be
closed.
The lagrangian of the SUSY Yang-Mills theory reads as follows
\cite{book}
\beq
L = \sqrt{g} \frac{1}{e^2} {\rm Tr}\; [\frac{1}{4} F_{\mu\nu} F^{\mu\nu}\;
+\; \bar{\lambda} i\dd \;\lambda\; +\;\frac{1}{2} D^2]
\label{lagrgauge}
\eeq
where $\bar{\lambda}$ and $\lambda$ stand for the right and left-handed Weyl
spinor correspondingly, $F_{\mu \nu}$ is the strength tensor for the
gauge field; $e^2$ stands for the gauge coupling constant,
$g$ is a determinant of the external metric tensor,
$\dd = \sigma^{\mu} D_{\mu} ,$  $D_{\nu} =\nabla_{\nu}
(\omega ,V)- iA_{\nu} ,$
and $\nabla_{\mu} = \partial_{\mu}
-i\omega^{ab}_{\mu}\sigma_{ab}/4 +V_{\mu} .$
This Lagrangian is invariant under a global supertransformation
generated by the right-handed
supergenerator $Q$ which corresponds to the constant spinor
$\bar{\epsilon}$ obeying eq.(2.3).
This supertransformation reads as follows
\beq
\delta A_{\mu}\;=\;(\bar{\epsilon} \sigma_a \lambda) e^a_{\mu} \; ,
\;\;\delta \lambda \;=\;0\;,
\label{transgauge}
\eeq
$$\delta\bar{\lambda}\;=\; D \bar{\epsilon} -\frac{1}{2}\;
F_{\mu \nu}\bar{\epsilon}
\sigma^{ab} \;e^{\mu}_a e^{\nu}_b,\;\;
\delta D\;=\; -\bar{\epsilon} i\dd \lambda .$$
By a direct calculation one can check that $Q$ is nilpotent, i.e.
$Q^2 =0$ and the Lagrangian is $Q$-exact
\beq
L\;=\;\sqrt{g} \frac{1}{2e^2}
\left\{ Q,\; \bar{\lambda} \left(
D\;+\;\frac{1}{2}\sigma^{\mu \nu} F_{\mu \nu}\right) \bar{\eta}\right\} ,
\label{Qlagrgauge}
\eeq
where a constant spinor $\bar{\eta}$ is linearly independent of
$\bar{\epsilon}$ and is normalized by $\bar{\epsilon}
\bar{\eta}=1 .$
\eq{Qlagrgauge} for the Lagrangian is equivalent to \eq{lagrgauge} up to
a term which is proportional to a topological charge of the gauge field.

Let us now consider a chiral supermultiplet in a representation $R$ of
the gauge group.
The components of a chiral supermultiplet are a complex, scalar field
$\phi ,$ a Weyl fermion $\psi$ and a complex scalar auxiliary field
$F.$
It is important however that the axial current of fermions which belong to the
supermultiplet of matter
be coupled to the twisting vector field $V_{\mu}$ with
an opposite charge to the one of the gaugino current.
This fact will allow us to cancel an anomaly of the total axial current
which is used for the twisting.
The absence of the anomaly for this current is a necessary \cn \ for an
interpretation of the twisted theory as a topological one (see below).

The Lagrangian for such a supermultiplet coupled to the
gauge supermultiplet in the presence of the external K\"ahler metric and
a vector (twisting) field $V_{\mu}$ reads
\beq
L = \sqrt{g} (D_{\mu} \bar{\phi} D^{\mu} \phi +
\bar{\psi} i\dd \psi +\bar{F}F + \bar{\phi} D\phi - 2\bar{\phi} \lambda
\psi - \bar{\psi} \bar{\lambda} \phi ).
\label{lagrmatt}
\eeq
This Lagrangian is invariant under the following SUSY transformation
\beq
\delta \phi =0, \;\;\; \delta\psi = -i\sigma_{\mu} \bar{\epsilon}D^{\mu}
\phi ,\;\;\; \delta F= -i\bar{\epsilon}\dd \psi + \bar{\lambda}
\bar{\epsilon} \phi,
\label{transmatt}
\eeq
$$\delta\bar{\psi}= \bar{\epsilon}\bar{\psi} ,\;\;\; \delta\bar{\psi}=
\bar{F} \bar{\epsilon} ,\;\;\; \delta\bar{F} =0.$$
Here the constant spinor $\bar{\epsilon}$ obeys \eq{defeps}.
The generator $Q$ of this transformation is nilpotent ($Q^2 =0$) and the
Lagrangian \rf{lagrmatt} is $Q$-exact
\beq
L = \{ Q, \bar{\eta} (-\bar{\psi} F +i \bar{\phi}\dd \psi -
\bar{\phi} \bar{\lambda} \phi) \} .
\eeq
Notice that the generator $Q$ of the supertransformations
\rf{transgauge}, \rf{transmatt}
depends on the external metric and the covariant
derivative acting to the spinors
includes an additional vector field $V_{\mu} ,$ i.e. $\dd =\sigma^{\mu}
(\partial_{\mu} -i\sigma_{ab} \omega^{ab}_{\mu} /4 +V_{\mu}
-iA_{\mu}).$
Now we want to interpret the generator $Q$ as a BRST charge in order to
define the theory as a topological one.
However the dependence of the BRST charge on external metric does not
allows us to do that since the variation of the action with respect to
the external metric is not $Q$-exact.
To avoid this difficulty
we shall change the spins of the quantum fields in order to remove
the metric from the definition of the generator $Q$ and to formulate
the Lagrangian in terms of usual covariant derivatives.
In this way we shall introduce new fundamental fields
which absorb some components of the metric tensor.

\subsection{Twisting of a supersymmetric theory}

It is convenient at this point to introduce a complex structure $J^{\mu}_{\nu}$
on $M$ associated to the spinors $\bar{\epsilon}$ and $\bar{\eta}$
\beq
J^{\mu}_{\nu} = (\bar{\eta} \sigma^{ab} \bar{\epsilon}) e^{\mu}_a
e_{b\nu} .
\eeq
We have
\beq
D_{\lambda} J^{\mu}_{\nu} =0 ,\;\;\; (J^2)^{\mu}_{\nu} = -
\delta^{\mu}_{\nu} ,
\eeq
where $D_{\lambda}$ is the untwisted covariant derivative on $M .$
With this complex structure one can define the holomorphic ($z^m ,$
$m=\bar{1},\bar{2}$) and
antiholomorphic ($\bar{z}^{\bar{m}} ,$ $\bar{m} =1,2$)
coordinates on $M$ so that
\beq
J^m_n z^n=iz^m,\;\;\; J^{\bar{m}}_{\bar{n}}
\bar{z}^{\bar{n}}=-i\bar{z}^{\bar{m}} .
\eeq
In this well adapted frame (see, e.g. \cite{green}) the complex
structure has the simplest form
\beq
J^m_n =i\delta^m_n ,\;\;\; J^{\bar{m}}_{\bar{n}} =
-i\delta^{\bar{m}}_{\bar{n}} ,\;\;\; J^{\bar{m}}_n =0,\;\;\;
J^m_{\bar{n}} =0
\eeq
and, hence,
\beq
J^{m\bm} = ig^{m\bm},
\eeq
while the other components of the metric tensor $g^{\mu\nu}$ vanish.

Let us now redefine the quantum fields.
After an appropriate redefinition \cite{ajoh} the gauge multiplet
contains a gauge field $A_{\mu} = (A_n ,A_{\bar{n}})$ (with the
holomorphic and anti-holomorphic components), a scalar fermion
$\bar{\lambda} = i\bar{\eta} \bar{\lambda}$ and a fermion $(0,2)$ form
$\bar{\lambda}_{\bar{m}\bar{n}}= (\bar{\eta} \sigma_{\bar{m}\bar{n}}
\bar{\eta}) (\bar{\epsilon} \bar{\lambda})$ (they both originate from the right
handed gaugino), a fermion $(1,0)$-form $\chi_n = \bar{\epsilon}
\sigma_n \lambda$ (it corresponds to the left handed gaugino), and an
auxiliary field $D' = iD + ig^{\bar{m} m} F_{\bar{m} m} ,$ where
$F_{\bar{m} m}$ is the strength tensor components of the gauge field.
Similarly for the chiral multiplet we have
a scalar complex bosonic field $\phi$ ($\bar{\phi}$),
a scalar fermion $\bar{\psi}$, the fermionic fields $\psi_{\bar{m}}$ and
$\bar{\psi}_{mn}$ which are $(0,1)$ and $(2,0)$ forms respectively, and
the auxiliary bosonic $(0,2)$ and $(2,0)$ forms $N_{\bar{m}\bar{n}}$ and
$\bar{N}_{mn} .$
The total twisted Lagrangian reads as follows \cite{ajoh}
\beq
L =\sqrt{g} \frac{1}{e^2} {\rm Tr} [ F^{\bar{m}\bar{n}} F_{\bar{m}\bar{n}}
+i \bar{\lambda}^{mn} D_m \chi_n -
\frac{1}{2} D'^2 + ig^{\bar{m} m} D' F_{\bar{m} m}
+i\bar{\lambda} D^m \chi_m] +
\label{totalL}
\eeq
$$+\sqrt{g} ( \bar{\phi} D^{\bar{m}} D_{\bar{m}} \phi +
\bar{\psi}^{\bar{m}\bar{n}} D_{\bar{m}} \psi_{\bar{n}} + \bar{\psi}
D^{\bar{m}} \psi_{\bar{m}} + N_{\bar{m}\bar{n}}
\bar{N}^{\bar{m}\bar{n}}-$$
$$-i\bh \c_n \p^n - i \bp \bl \h  +\frac{1}{4} \bp_{mn}
\bl^{mn} \h + i \bh D' \h ) =$$
$$= \sqrt{g}
\{ Q, \frac{1}{e^2} {\rm Tr}[-i\bar{\lambda}  g^{\bar{m} m} F_{\bar{m} m}
+\frac{1}{2} D' \bar{\lambda} -
\frac{i}{2} \bar{\lambda}_{\bar{m}\bar{n}}
F^{\bar{m}\bar{n}}]\} +$$
$$ +\sqrt{g}
\{ Q, -\frac{1}{2} \bar{\psi}^{\bar{m}\bar{n}}N_{\bar{m}\bar{n}} +
\bar{\phi} D^{\bar{m}} \psi_{\bar{m}} -i\bar{\phi}
\bar{\lambda} \phi \} .$$
Here $Q$ is the scalar nilpotent generator of BRST transformations
which for the gauge multiplet read
\beq
\delta A_n = \chi_n,\;\;\delta A_{\bar{n}} =0,\;\;
\delta \chi_n =0,
\eeq
$$\delta \bar{\lambda} = - D',\;\;
\delta \bar{\lambda}_{\bar{m} \bar{n}} =
2i F_{\bar{m} \bar{n}},\;\;
\delta D' = 0,$$
while for the chiral multiplet we have
\beq
\delta \phi =0,\;\;\; \delta \psi_{\bar{m}} = D_{\bar{m}}
\phi ,\;\;\; \delta N_{\bar{m}\bar{n}}= D_{\bar{m}}
\psi_{\bar{n}} - D_{\bar{n}}\psi_{\bar{m}} +\frac{1}{2}
\bar{\lambda}_{\bar{m}\bar{n}} \phi ,
\eeq
$$\delta\bar{\phi} =\bar{\psi} ,\;\;\;
\delta \bar{\psi} =0 ,\;\;\;
\delta \bar{\psi}^{\bar{m}\bar{n}} = -2 \bar{N}^{\bar{m}\bar{n}} ,\;\;\;
\delta \bar{N}^{\bar{m}\bar{n}} =0 .$$
It is easy to see that the generator of these transformations does not
depend on the external metric.

One can see that all the fields are combined into different
supermultiplets.
Indeed in the gauge sector we have the following supermultiplets:
($A_n ,\; \chi_n$), ($\bar{\lambda},\; D'$) and
($\bar{\lambda}_{\bar{m} \bar{n}},\; F_{\bar{m} \bar{n}}$).
In the matter sector the multiplets are $(\phi ,\psi_{\bar{m}} ,
N_{\bar{m}\bar{n}}) ,$
$(\bar{\phi} ,\bar{\psi}) ,$ and $(\bar{\psi}_{mn} , \bar{N}_{mn} ).$

By fixing the ghost number of the BRST charge to be 1 we have
the following dimensions ($d$) and
ghost numbers ($G$) for the fields:
\beq
(d,G)(A_n) = (d,G)(A_{\bar{n}}) = (1,0), \;\;
(d,G)(\chi_n) = (1,1)\;\;
\eeq
$$(d,G)(\bar{\lambda}) = (d,G)(\bar{\lambda}_{\bar{m} \bar{n}})=
(2,-1) ,\;\;
(d,G)(D') = (2,0).$$
and
\beq
(d,G) (\phi) = (0,2),\;\;\; (d,G)(\bar{\phi})=(2,-2),\;\;\;
(d,G)(\psi_{\bar{m}}) =(1,1),
\eeq
$$(d,G)(\bar{\psi}) = (d,G)(\bar{\psi}_{mn}) =(2,-1),\;\;\;
(d,G)(N_{\bar{m}\bar{n}})=(2,0),\;\;\; (d,G)(\bar{N}_{mn})=(2,0) .$$
Notice also that for the case  of hyperk\"ahler manifold $M$ the
holonomy group is reduced to the $SU(2)$ one and twisted and untwisted
theories coincide \cite{ajoh,witten4}.

\subsection{Twisted self-interacting chiral multiplet}

Let us now consider the theory without gauge interactions.
The Lagrangian for a twisted
free chiral multiplet reads as follows
\beq
L_0 = \sqrt{g} ( \bar{\phi} D^{\bar{m}} D_{\bar{m}} \phi +
\bar{\psi}^{\bar{m}\bar{n}} D_{\bar{m}} \psi_{\bar{n}} + \bar{\psi}
D^{\bar{m}} \psi_{\bar{m}} + N_{\bar{m}\bar{n}}
\bar{N}^{\bar{m}\bar{n}}) =
\label{freeCH}
\eeq
$$=\sqrt{g}
\{ Q, -\frac{1}{2} \bar{\psi}^{\bar{m}\bar{n}}N_{\bar{m}\bar{n}} +
\bar{\phi} D^{\bar{m}} \psi_{\bar{m}}\} .$$
Here $Q$ is the scalar nilpotent generator of BRST transformations
\beq
\delta \phi =0,\;\;\; \delta\psi_{\bar{m}} = \partial_{\bar{m}} \phi ,\;\;\;
\delta N_{\bar{m}\bar{n}} = D_{\bar{m}} \psi_{\bar{n}} - D_{\bar{n}}
\psi_{\bar{m}} ,
\label{freeQ}
\eeq
$$\delta \bar{\phi} =\bar{\psi} ,\;\;\; \delta \bar{\psi} =0,\;\;\;
\delta \bar{\psi}_{mn} = -2 \bar{N}_{mn} ,\;\;\; \delta \bar{N}_{mn} =0
.$$
If there is a nontrivial holomorphic $(2,0)$ form $E_{mn}$ on $M$
(i.e. $H^{2,0} (M) \neq 0$) then it is
possible to introduce a superpotential $W(x)$ which induces masses
and self-interactions for the quantum fields.
The corresponding interacting Lagrangian reads
\beq
L_{int} = \sqrt{g}
[E^{\bar{m}\bar{n}} (\psi_{\bar{m}} \psi_{\bar{n}} W'' (\phi) +
N_{\bar{m}\bar{n}} W' (\phi)) -
S^{mn} (\bar{\psi}_{mn}\bar{\psi} W'' (\bar{\phi}) + 2 \bar{N}_{mn} W'
(\bar{\phi}))]=
\label{selfint}
\eeq
$$= \sqrt{g} [E^{\bar{m}\bar{n}} (\psi_{\bar{m}} \psi_{\bar{n}} W'' (\phi) +
N_{\bar{m}\bar{n}} W' (\phi)) +\{ Q, S^{mn} \bar{\psi}_{mn} W'
(\bar{\phi}) \} ] ,$$
where $S_{\bar{m}\bar{n}}$ is an arbitrary non-singular (0,2) form on $M
.$
It is easy to see that
\beq
\{ Q, \int_M L_{int} \} =0,
\eeq
and hence the total action of the theory with a superpotential is
$Q$-closed (but not $Q$-exact due to the interaction terms).

Notice that a superpotential can also be introduced into the heterotic
topological gauge theory.
In this case the action is $Q$-closed but not
$Q$-exact.
Moreover as we shall see below the heterotic topological gauge theory has
a moduli space corresponding to deformations of the Lagrangian by a
superpotential.

The \cn \ of renormalizability of four dimensional theory implies that
$W(x)$ should be a polynomial of a degree not higher than 3.
In the case of the superpotential of a degree 2 a Majorana mass
is induced for the
chiral supermultiplet, while for the cubic superpotential the $|\phi|^4$ +
Yukawa interactions are induced.
As it is well known \cite{book} the superpotential in SUSY theories is
non-renormalizable \footnote{Notice however that some subtleties can appear due
to infrared effects \cite{IR}.}.
The only renormalization of the action comes from $D$-terms \cite{book}.
This fact was very important in the analysis of 2D N=2 SUSY theories
\cite{Ma,Va,LVW,howe,cec1,cec2,ring,WiLG}, where
the $N=2$ superconformal theories were
associated to quasihomogeneous superpotentials.
As we demonstrate below the non-renormalizability of a superpotential
is also important for a conformal structure that appears in the
heterotic topological theory.
For simplicity we shall also consider the case of $W(x)=
\lambda x^{n+1}/(n+1) ,$ where $\lambda$ is a coupling constant,
and formally we shall consider all positive integer values of $n.$

\subsection{Physical \op s}

We can now define the physical \op s as classes of cohomology of the
BRST \op \ $Q.$
The local observable for the sector of the gauge multiplet
becomes a (2,0) form (of dimension 2)
\beq
O^{(0)}_{mn} = {\rm Tr} \chi_m \chi_n .
\eeq
It is to be noticed that the situation here is different from the
ordinary topological theories where the local observables are zero-forms;
non-zero forms should usually be integrated over closed cycles
to get non-local observables (in the case of highest forms one gets
moduli of the topological theory).
The difference here is due to the splitting of four coordinates
into holomorphic and anti-holomorphic ones, so that the (2,0) form
is effectively a scalar with respect to anti-holomorphic derivatives.
We have
\beq
\partial_{\bar{k}} {\rm Tr} \chi_m \chi_n = \{ Q,...\} .
\eeq
It follows from this equation that the physical correlators under
an insertion
of this operator are holomorphic with respect to its coordinate.
This is of course quite similar to the left-moving nature of the
cohomology in 2D half-twisted theories \cite{WiLG}.

One can also construct the non-local observables using the descent
procedure \cite{witten}.
We have
\beq
\partial_{\bar{k}} O^{(0)}_{mn} =
\{ Q, H^{(1)}_{mn, \bar{k}}\},
\label{deq}
\eeq
$$\partial_{[\bar{p}}  H^{(1)}_{mn, \bar{k}]} =
\{ Q , H^{(2)}_{mn, \bar{k}\bar{p}}\}$$
where
\beq
H^{(1)}_{mn, \bar{k}}=
{\rm Tr} (F_{\bar{k} m} \chi_n -F_{\bar{k} n} \chi_m ) ,
\eeq
$$H^{(2)}_{mn, \bar{k}\bar{p}} =
\frac{1}{2} {\rm Tr} (F_{\bar{k} m} F_{\bar{p} n}-
F_{\bar{k} n} F_{\bar{p} m} + F_{\bar{p}\bar{k}}
F_{m n}) +
\frac{i}{4} {\rm Tr} (\partial_m \bar{\lambda}_{\bar{p} \bar{k}}
\chi_n - \partial_n \bar{\lambda}_{\bar{p} \bar{k}}
\chi_m).$$
The \op \ $H^{(2)}_{mn, \bar{k}\bar{p}}$ is obviously
the density of the topological charge of the gauge field up to
an exact form.
These relations allow us to construct the following
$Q$-closed non-local observables
\beq
O^{(1)} =\int \bar{\omega} \wedge H^{(1)},
\;\;
O^{(2)} = \int  H^{(2)},
\eeq
where $\omega$ is a closed (0,1) form.
Here we used that the forms $H^{(1)}$ and
$H^{(2)}$ are $Q$-closed up to exact differential
forms according to eqs.~\rf{deq}.

The local \op s in the matter sector are given by the same
gauge invariant functions of the (dimensionless) scalar field $\phi$ as in the
supersymmetric version of the theory (because this scalar field has zero
axial charge and therefore does not change its spin under the twisting)
which correspond to flat directions of the classical
moduli space of vacua \cite{seiberg}.
If the cohomology space $H^{2,0} (M) \neq 0$ (i.e. there are holomorphic
$(2,0)$ forms on $M$)
then it is also possible to construct non-local operators.
Let the matter supermultiplets transform as a reducible representation
$R$ of the gauge group, where each irreducible representation
$R_i$ enters with a repetition $n_i .$
In particular let us consider two such
irreducible representations $R_i$ and $R_j$ for which the
tensor product $R_i \otimes R_j$ contains a singlet.
One can construct the following bilinear BRST-invariant \op \
\beq
O_{matter} = \sum_{IJ} a_{IJ}
\int_M E \wedge <\psi^I \wedge \psi^J + \frac{1}{2} N^I \phi^J
+ \frac{1}{2} \phi^I N^J > .
\label{Omatter}
\eeq
Here $<...>$ stands for a gauge invariant pairing of the fields
(its definition depends on a representation of the gauge group for the
fields of matter),
the indices $I$ and $J$ stand for copies of the irreducible
representations $R_i$ and $R_j$ of the matter fields,
and $E$ is a holomorphic (2,0) form on $M .$
$a_{IJ}$ is a constant matrix
(a choice of it depends on a representation of the fields of matter).
For example for the case when the multiplet of matter is in an adjoint
representation of the gauge group (this example corresponds to the
twisted $N=2$ supersymmetric Yang-Mills theory)
the matrix $a_{IJ}$ is just 1.
For the case of the $SU(2)$ gauge group where the fields of matter
fall in four copies
of a spinor representation, $a_{IJ}$ is an antisymmetric $4\times 4$
matrix.
The \op \ \eq{Omatter} is actually a twisted version of an $F$ term for
a bilinear combination of the chiral superfields \cite{book}.
This \op \
corresponds to mass deformations of the theory (see \eq{selfint}).

Similarly one can write down a BRST closed \op \ which is an integral of
a $(2,2)$ form and contains both the fields of the gauge and matter
sectors.
We have
\beq
O_{mix} =\sum_{IJ} a_{IJ}
\int_M [H^{(1)} \wedge <\h^I \p^J +\p^I \h^J> +H^{(2)}
<\h^I\h^J> +
\label{Omix}
\eeq
$$+ O^{(0)} \wedge <\p^I \wedge\p^J +\frac{1}{2}\h^I N^J +
\frac{1}{2} N^I \h^J >] .$$
This \op \ exists even for the manifolds with $H^{2,0} (M) = 0.$
One can of course construct more complicated physical \op s.

It is worth noticing that we could use the vector field $-V_{\mu}$ for a
twisting of the theory on a K\"ahler manifold.
Such a modification of the model corresponds to a change $\epsilon ,\eta
\to \eta ,\epsilon .$
The local \op s and their correlators in this mirror model
are antiholomorphic up to BRST exact \op s (for example, $\partial_n \phi =
\{Q , ...\}$).

Notice also that for the case of a hyperk\"ahler manifold
there is also another BRST charge since in this case the holonomy
group is $SU(2)$.
This BRST charge corresponds to a different right-handed
supergenerator with a parameter which is
proportional to the spinor $\bar{\eta}$ defined in section 2.1.
The holomorphic derivatives of the local physical \op s of the matter sector
turn out to be exact with respect to this second BRST charge.
Therefore the correlators of the local physical \op s
do not depend on all coordinates.

\subsection{Anomalies}

In this section we consider the problem of the quantum
anomaly for the BRST symmetry in the heterotic topological gauge
theories.

Let us consider such a theory without a superpotential for the matter
fields.
The Lagrangian of its twisted version is BRST exact at the classical level.
Due to the $Q$-exactness of the Lagrangian the metric is expected to
decouple from the physical correlators.
However as it is shown in ref.\cite{ajoh} on a curved
K\"ahler manifold an anomaly at the quantum level can
prevent the decoupling of the metric.
This anomaly appears because in the case of an arbitrary representation
of the matter multiplet the fermionic current which is
coupled to the twisting vector field $V_{\mu}$ is anomalous.

This anomaly can be easily computed at the one-loop level \cite{ajoh}.
Let us consider a variation of the effective action for external
gravitational and gauge fields (we define the effective action by
$S_{eff} = -\log \; Z,$ where $Z$ is the partition function).
This one-loop effective action is given by a ratio of the different
Laplace \op s on $M$ in the external gravitational and gauge fields
\cite{ajoh}.
The infrared contributions to the effective action appear due to the zero
modes of the Laplace \op s and are not related to the BRST anomaly that we
consider (this anomaly is related to the necessary ultraviolet
regularization of the theory).
The ultraviolet contribution to the variation of the effective action
can be represented as follows
\beq
\delta S_{gr} + V_{mix} [A_m ,A_{\bar{m}}, g_{m\bar{m}}] .
\label{vartot}
\eeq
Here $\delta S_{gr} (g)$ does not depend on the gauge field.
The term $V_{mix} [A_m ,A_{\bar{m}}, g_{m\bar{m}}]$
is a local functional of the external gauge field
because it is determined by ultraviolet contributions.

Let us consider the purely gravitational part of the variation
in \eq{vartot}.
It is obvious that $S_{gr}$ is proportional to the dimension of
the representation of the fields over which we integrate in the path integral.
The contribution from the gauge sector is proportional to the
dimension of the gauge group ${\rm
dim}\; G ,$ while that from the matter sector is proportional to $-{\rm
dim} \; R$ \cite{ajoh} (${\rm dim}\; R$ stands for the dimension of the
representation of the matter multiplet).
The difference in sign comes from the different statistics of vector
fields in the gauge and in the matter multiplets.
If we integrate the gravitational part of the
variation of the effective action and
normalize the path integral that divides it by the same path integral
without any external gauge field then we get the following purely
gravitational factor for the partition function $Z$
\beq
\left( \frac{T_2 (\zeta ,M)}{T_2 (0,M)} \right)^{{\rm dim}\; R -
{\rm dim}\; G} ,
\eeq
where $T_2 (\zeta ,M)$ is the Ray-Singer torsion \cite{ray}
($\zeta$ stands for a flat connection on $M$ associated with a
particular
representation of the fundamental group on $M$) and $T_2 (0,M)$ is an
ultraviolet contribution to the Ray-Singer torsion.
It can be shown that
this factor depends only on the K\"ahler class of the metric \cite{ray}.

It is worth emphasizing that the gravitational anomaly is $Q$-closed
since the external metric is BRST invariant.

Let us now consider the second part of the variation in \eq{vartot}.
Since it is a local functional of the gauge field
one can determine it by a
calculation in the case of a gauge field for which the Laplace operators
have no zero modes, in particular for small values of the gauge field.
It is useful at this point to compare the present situation with
the supersymmetric version of the theory (with the twisted spin
connection).
Actually there are two sorts of anomalies.
The first one is a conformal anomaly which is proportional
to $\int {\rm Tr} ^* F \wedge F \log g$ while the second anomaly appears
due to the coupling of the anomalous axial current to the external vector
field $V_{\mu} .$
The axial anomaly for the gaugino current
gives the following term in the effective action $\log Z$
\beq
\frac{1}{16\pi^2} \int_M {\rm Tr}_{Ad} F \wedge F \frac{1}{\Delta}
D_{\mu} V_{\mu}= \frac{1}{32\pi^2}
\int_M {\rm Tr}_{Ad} F\wedge F \log e/\bar{e} .
\label{axial}
\eeq
where we used the fact
that locally the vector field $V$ is a total derivative \cite{ajoh}.
Here ${\rm Tr}_{Ad}$ stands for the trace taken in an adjoint representation.
The contribution from the matter sector to the axial anomaly is given by
the same expression \rf{axial} with an opposite sign and
with the trace ${\rm Tr}_R$ taken in the
representation $R$ of the gauge group.

For the twisted theory formulated in terms of sections of
holomorphic and anti-holomorphic vector bundles over $M$ the
anomaly can be determined by a direct calculation \cite{ajoh}.
We get the following expression for $V_{mix}$
\beq
-\frac{1}{16\pi^2} \int_M {\rm Tr}_{Ad}
[g^{n\bar{m}}\delta g_{k\bar{m}}(-F^{k\bar{l}}
F_{\bar{l} n} + F_p^p F^k_n) -
g^{m\bar{m}} \delta g_{m\bar{m}} (-F^{k\bar{l}}
F_{\bar{l} k} + (F_p^p)^2)] +
\eeq
$$+\frac{1}{16\pi^2} \int_M {\rm Tr}_R
[g^{n\bar{m}}\delta g_{k\bar{m}}(-F^{k\bar{l}}
F_{\bar{l} n} + F_p^p F^k_n) -
g^{m\bar{m}} \delta g_{m\bar{m}} (-F^{k\bar{l}}
F_{\bar{l} k} + (F_p^p)^2)] .$$
Let the variation of the metric be purely K\"ahlerian, i.e.
$\delta g_{m\bar{m}} = \partial_m \omega_{\bar{m}} + \partial_{\bar{m}}
\omega_m$ where $\omega_m$ and $\omega_{\bar{m}}$ are (1,0) and (0,1)
forms so that the K\"ahler forms $J$ and $J+ d\omega$ belong to the same
cohomology class in $H^2 (M,R) .$
Then this variation can be integrated and we get the following expression
for an anomalous contribution to the effective action $\log Z$
\beq
-\frac{1}{32\pi^2} \int_M  {\rm Tr}_{Ad} F \wedge F \log g +
\label{twistanom}
\eeq
$$+\frac{1}{32\pi^2} \int_M  {\rm Tr}_R F \wedge F \log g.$$
The anomaly in this form is similar to that in \eq{axial} but it
takes into account the change of the path integral
measure when we translate the supersymmetric theory into the twisted one.

Actually it is easy to check that
if we take into account the contributions to
the effective action which contain the gaugino fields then
the total anomaly in the effective action is proportional to
\beq
\int_M  H^{(2)} \log g ,
\label{totanom}
\eeq
where the \op \ $H^{(2)}$ is defined as in the previous section.
This anomaly is obviously not BRST invariant and breaks the BRST invariance
of the theory at the quantum level.

For an
external anti-instanton field this term in the effective action obviously
mixes the dependence of the metric and of the moduli of instanton.
Hence in order to get a topological theory we have to cancel this anomaly by
contributions of matter in an appropriate representation of the gauge group.

Finally for the total contribution of the non-zero modes to the
partition function we get
\beq
Z = \frac{\hat{T}_2 (M,\zeta ,0,R)}{\hat{T}_2 (M, \zeta ,A,R)}
\frac{\hat{T}_2 (M,\zeta ,A,Ad)}{\hat{T}_2 (M, \zeta ,0,Ad)} ,
\eeq
where $\hat{T}_2 (M, \zeta ,A,R)$ stands for a generalized Ray-Singer
torsion in an external non-abelian gauge field $A$ and in the presence of a
flat connection $\zeta$ associated with a particular representation of
the fundamental group of $M$ \cite{ajoh}.

{}From \eq{twistanom}
we can see that the mixed anomaly $V_{mix} [A_m ,A_{\bm} ,g_{m\bm}]$
is cancelled if
\beq
C_2 (G) -T(R) = C_2 (G) -\sum_i T(R_i)=0.
\label{conMix}
\eeq
Here $R=\sum_i R_i ,$ $R_i$ are irreducible representations of the gauge
group; $C_2 (G)$ stands for the Casimir \op \
(${\rm Tr}_{Ad}
t^a t^b = C_2(G) \delta^{ab}$) and $T(R_i)$ is the Dynkin index of
an irreducible representation $R_i$ of the group $G$ (${\rm Tr}_R
t^a t^b = T(R) \delta^{ab} ,$ $t^a$ and $t^b$ stand for the generators of the
gauge group).
The condition for the cancellation of the gravitational anomaly (the
Ray-Singer torsion) reads
\beq
{\rm dim} \; G - \sum_i {\rm dim} \; R_i  = 0,
\label{conGrav}
\eeq
Where ${\rm dim}\; G$ and ${\rm dim}\; R_i$ stand for the dimensions of the
adjoint and $R_i$ representations of the gauge group respectively.
A \cn \ \rf{conMix} has been analyzed (for a different problem) in
ref.\cite{koh}.
{}From their analysis one can easily extract that if the \cn \ \rf{conMix} is
fulfilled then
\beq
\sum_i {\rm dim} \; R_i - {\rm dim}\; G  \geq 0 .
\eeq
The equality in the above equation is reached only for matter in the
adjoint representation of the gauge group, i.e. for the twisted $N=2$
Yang-Mills theory \cite{witten}.
In the Appendix we list the representations of all classical groups
$A_n ,$ $B_n ,$ $C_n ,$ $D_n ,$ $E_6 ,$ $E_7 ,$ $E_8 ,$ $F_4 ,$ $G_2$
which obey \eq{conMix} and give the corresponding values of
$\sum_i {\rm dim} \; R_i - {\rm dim}\; G.$

{}From now on we shall consider only theories where the mixed anomaly
is cancelled
\footnote{The mixed anomaly is clearly absent in the case of a twisted
model of a chiral supermultiplet without gauge interactions.}.
Notice that the condition for a cancellation of the mixed anomaly does not
imply that the variation of the external metric is purely K\"ahlerian,
i.e. does not change the K\"ahler class of the metric.
It is easy to see that the mixed anomaly is cancelled provided that
the \cn \ \rf{conMix} is satisfied for an arbitrary variation of the
K\"ahler external metric.

As to the purely gravitational part of the (normalized) partition
function it can depend only on the K\"ahler class of the metric.
Thus the whole physical correlator remains unmodified under
smooth variations of the external K\"ahler metric
provided that they do not change its K\"ahler class.
Therefore we do not impose the \cn \ \rf{conGrav}.
Moreover as we shall see below the gravitational part of the
non-normalized partition function has an interesting interpretation as a
conformal anomaly of an embedded conformal theory
\footnote{There exists one more restriction to the matter sector:
the theory should not have (both local and global) gauge anomalies.}.

Our analysis was restricted above to the one-loop level.
It is important to understand if the above arguments can be extended to
the multiloop level.
The absence of a mixed anomaly at the one-loop level is sufficient for
the vanishing of multiloop contributions to the mixed anomaly.
This is because the anomaly originates essentially from the axial
anomaly of the fermionic current coupled to the external vector field
$V_{\mu} .$
The multiloop corrections to the axial anomaly at the level of matrix
elements can appear only due to a rescattering of gluons \cite{anj}
provided the one-loop anomaly is not cancelled.
Thus the mixed anomaly is absent provided that the \cn \ \rf{conMix} is
satisfied.

If the mixed anomaly is cancelled the BRST invariance is not broken by
the quantum corrections
since the external metric and, hence, the effective action is
$Q$-invariant (the multiloop corrections to the gravitational anomaly are
discussed below).
It is also easy to see that the variation of this Lagrangian both in
the gauge coupling constant is $Q$-exact.
Therefore the physical correlators do not depend on the gauge coupling constant
and the theory formally allows for a localization near the solutions of
the classical equations of motion similar to usual topological theories
\cite{witten}.
In such theory one can try to calculate the physical correlators in the
limit of weak coupling $e^2\to 0.$
However there is a subtlety in such an approach.
The point is that the theory under consideration is, in general, strongly
interacting and the coupling constant $e^2$ may become of order 1 due to
infrared contributions since it depends on a renormalization scale.
In order to keep the coupling constant small one has to ``freeze'' it by
a Higgs mechanism or by suppressing the infrared effects
via the introduction of a mass gap into the theory.
Indeed Witten \cite{witten4} recently calculated Donaldson invariants on
four-dimensional K\"ahler manifolds in the infrared limit by introducing
such a mass gap into the topological Yang-Mills theory.
The option to introduce a mass gap is technically very important
but not, strictly speaking, necessary for the physical
correlators to be topologically invariant.
In fact this has been demonstrated by formulas \cite{witten4} for
the Donaldson invariants which allow for their expansion
into series in powers of the mass parameter.

Actually the infrared problem which appears in the topological theories
is a counterpart of those in the supersymmetric theories on flat
space-time \cite{vashi,IR}.
The so-called non-renormalizability theorems are based essentially on
the existence of an infrared cut-off in the theory.
Such an infrared cut off can be provided by a mass gap (as in the case of
the non-renormalizability theorem for a superpotential) or by an
appropriate external field (as in the case of
the non-renormalizability theorem for an effective action in an external
instanton field) \cite{grisaru,vainshtein}.

In next subsection we consider the problem of renormalization of the
gravitational anomaly.

\subsection{Induced gravity and renormalizations}

The variation of the Lagrangian in the external metric is not
strictly speaking $Q$-exact at the quantum level
since the effective action is $Q$-closed but not $Q$-exact.
However such a dependence of the effective action (and of the physical
correlators) on the external metric is factorized out at the one-loop
level.
Therefore it does not spoil the BRST invariance of the theory provided
the multiloop corrections to the gravitational anomaly vanish.

Actually in four dimensional quantum field theories coupled to
the external gravitational the multiloop contributions to the induced
gravitational effective action are not usually vanishing \cite{trace}.
We shall argue however that in the heterotic topological theories
the gravitational anomaly does not acquire any multiloop corrections.
The absence of multiloop corrections maintains the BRST invariance of
the theory at the quantum level
because otherwise the ultraviolet logarithms that appear due to
interactions of quantum fields could induce mixed
gravity-gauge field terms in the effective action (such terms would
spoil the BRST invariance of the theory).
In what follows
we give the arguments which are valid for any heterotic topological
theory.

{}From the point of view of supersymmetry the gravitational anomaly
originates in the anomaly of the axial fermionic
current which is
coupled to the external vector field $V_{\mu}$ (see section 2).
This axial current has an anomaly that depends only on the classical
external gravitational fields.
The anomalous dimension of the current (which could be
responsible for multiloop corrections to the axial anomaly at the level
of matrix elements) is due to a rescattering of the fields which enter
the anomaly.
In our case the gravitational field is the external one and does not
induce any diagrams with virtual gravitons.
Therefore the current does not have any anomalous dimension and hence
the anomaly is not renormalizable.

A different argument is based on the background
superfield formalism of supergravity \cite{marcgrav}.
As it has been demonstrated in ref.\cite{marcgrav} a superfield Feynmann
diagram for multiloop corrections to the effective action
can be represented as an integral over four Grassmann variables
with an integrand which is a local expression with respect to the
Grassmann coordinates.
When we formulate our theory in terms of the
usual fields (with unmodified spins) the quantum supermultiplets are
coupled to a reduced external gravitational superfield
that is invariant under one of four supercharges.
This means that the integrands in the superfield diagrams in such a
background do not depend on one of the Grassmann variables.
Therefore such multiloop corrections to the effective action in this
special supergravity background vanish due to an integration over
Grassmann coordinates.
It is straightforward to reformulate this conclusion in terms of
twisted fields.
We thus see that there is no multiloop corrections to the gravitational
anomaly \rf{effGRt},\rf{effGRr}.
This argument is an extension of the usual theorems of
non-renormalizability for a superpotential and of the effective action
in the instanton background \cite{grisaru,vainshtein} to the case of
a special supergravitational background.

We shall demonstrate by an explicit calculation
the vanishing of the 2-loop correction to the effective gravitational action
in the model of a single chiral supermultiplet with a superpotential
$W (x)$ (we assume that $H^{2,0} (M) \neq 0$).
For simplicity we take $W(x)= x^3/12 .$
In this model the interaction is described by the following terms in the
Lagrangian (see \eq{selfint})
\beq
L_{int} = \sqrt{g}
[E^{\bar{1}\bar{2}} (\p_{\bar{1}}\p_{\bar{2}} \h + \frac{1}{2}
N_{\bar{1}\bar{2}} \h^2) - S^{12} (\bp_{12} \bp \bh + \bar{N}_{12}
\bh^2)] ,
\label{int}
\eeq
where $E_{12}$ is a holomorphic $(2,0)$ form and $S_{\bar{1}\bar{2}}$
stands for a non-singular $(0,2)$ form.
Actually the vanishing of multiloop corrections to the gravitational
effective action formally follows from the $Q$-exactness of the terms
in the Lagrangian \rf{int} which are proportional to
$S_{\bar{1}\bar{2}} .$
For the correction to the effective action we have
\beq
\int_M d^4 x \sqrt{g(x)}\int_M d^4 y \sqrt{g(y)}S^{12}(y) E_{12}(x)
\eeq
$$\left\{- <N^{12} (x) N_{12} (y)> <\h (x) \bh (y)> <\bh (y) \h (x)> +
\right.$$
$$ +<\h (x) \bh (y)> \left[ <\p_{\bar{1}} (x)\bp_{12} (y)>
\cdot <\p_{\bar{2}} (x)\bp (y) > -\right. $$
$$\left. \left.
-<\p_{\bar{2}}(x)\bp_{12}(y)>\cdot <\p_{\bar{1}}(x)\bp (y)>
\right] \right\} .$$
By substituting in the free propagators given in Appendix (\eq{freeprop})
we have
\beq
\int_M d^4 x \sqrt{g(x)}\int_M d^4 y g^{-1} (y)
S^{12}(y) E_{12}(x) \left\{
-\frac{1}{2} \delta^4 (x-y) \left(\frac{1}{\Delta_{00}} \delta^4 (x-y)
\right)^2 -\right.
\label{corr}
\eeq
$$- \left(\frac{1}{\Delta_{00}} \delta^4 (x-y)\right)
\left[ \left( D_{\bar{2}}\frac{1}{\Delta_{00}} \delta^4 (x-y)\right)
\cdot \left(\left(\frac{1}{\Delta_{0,1}}\right)_{\bar{1} 1} D^{\bar{2}}
\delta^4 (x-y) \right) + \right.$$
$$\left. \left. + \left( D_{\bar{1}}\frac{1}{\Delta_{00}} \delta^4 (x-y)\right)
\cdot \left(\left(\frac{1}{\Delta_{0,1}}\right)_{\bar{2} 2} D^{\bar{1}}
\delta^4 (x-y) \right) \right] \right\} .$$
By using
$$\left(\frac{1}{\Delta_{00}} \delta^4 (x-y)\right)
D_{\bar{n}}\left(\frac{1}{\Delta_{00}} \delta^4 (x-y)\right)=
\frac{1}{2} D_{\bar{n}}\left(\frac{1}{\Delta_{00}} \delta^4
(x-y)\right)^2$$
and taking into account a holomorphicity of the form $E_{12}$ one can
integrate by parts in the second line of \eq{corr}.
Thus we get
\beq
\int_M d^4 x \sqrt{g(x)} \int_M d^4 y g^{-1} (y) S^{12}(y) E_{12}(x) \left\{
-\frac{1}{2} \delta^4 (x-y) \left(\frac{1}{\Delta_{00}} \delta^4 (x-y)
\right)^2 +\right.
\eeq
$$\left. + \frac{1}{2} \left( \frac{1}{\Delta_{00}} \delta^4
(x-y)\right)^2
\left( D_{\bar{2}}\left(\frac{1}{\Delta_{0,1}}\right)_{\bar{1} 1}
D^{\bar{2}}
\delta^4 (x-y) +
D_{\bar{1}}\left(\frac{1}{\Delta_{0,1}}\right)_{\bar{2} 2} D^{\bar{1}}
\delta^4 (x-y) \right)\right\} =0 .$$
The latter equality follows from the following identity
\beq
D_{\bar{2}}\left(
\frac{1}{\Delta_{0,1}}\right)_{\bar{1} 1} D^{\bar{2}}
\delta^4 (x-y) +
D_{\bar{1}}\left( \frac{1}{\Delta_{0,1}}\right)_{\bar{2} 2} D^{\bar{1}}
\delta^4 (x-y) = \delta^4 (x-y) ,
\eeq
which is a manifestation of the supersymmetric Ward identity
(for the free fields)
\beq
2<N_{\bar{1}\bar{2}} (x) \bar{N}_{12} (y)> -
<(D_{\bar{1}} \p_{\bar{2}}-D_{\bar{2}} \p_{\bar{1}}) (x) \bp_{12} (y)>
= <\{ Q, N_{\bar{1}\bar{2}} (x) \bp_{12} (y) \} > =0.
\eeq
This calculation is of course formal since we did not introduce any
ultraviolet regularization.
But it demonstrates the origin of the cancellation of multiloop
corrections.

Notice that one could expect the BRST cohomology not to be well
defined in the presence of the gravitational anomaly since it is BRST
closed but not exact.
However
because the gravitational anomaly is not renormalizable one can cancel
it, for example, by adding to the model
$({\rm dim} R- {\rm dim} G)$ free $U(1)$ gauge supermultiplets.
This means that the BRST cohomology is defined and not spoiled by this
gravitational anomaly.

The gravitational anomaly for heterotic topological theories
has a natural interpretation
as a conformal anomaly of the Virasoro algebra which appears in the $Q$
cohomology (see below).
Indeed the form of the corresponding term in the effective action can be
extracted from the 4D conformal anomaly in the trace of the
energy-momentum tensor (see, e.g. \cite{duff})
which in our case reads
\beq
<\theta_{\mu}^{\mu}> = ({\rm dim} R - {\rm dim} G)
\frac{1}{3\cdot 128\pi^2} \; ^*R_{\mu\nu\lambda\rho}^*
R^{\mu\nu\lambda\rho} ,
\eeq
where $R_{\mu\nu\lambda\rho}$ is the Riemann tensor.
The variation of the effective action $S_{eff}$ with respect to the metric
$g_{\mu\nu}$ reads
\beq
\delta S_{eff} = ({\rm dim} R - {\rm dim} G)
\frac{1}{6\cdot 128\pi^2}\int_M d^4 x \sqrt{g}
 ^*R_{\mu\nu\lambda\rho}^* R^{\mu\nu\lambda\rho} \delta \log g ,
\eeq
and hence for the induced gravitational action we get
\beq
({\rm dim} G - {\rm dim} R)
\frac{1}{3\cdot 128\pi^2}\int_M d^4 x \sqrt{g}\;
 ^*R_{\mu\nu\lambda\rho}^* R^{\mu\nu\lambda\rho} \frac{1}{\Delta} R ,
\label{effGRt}
\eeq
where $\Delta$ is the Laplace \op ,
and $R= g^{m\bar{m}} \partial_m \partial_{\bar{m}} \log \det g_{m\bm}$
(see, e.g. \cite{green}).

For the case $M= \Sigma_1 \times \Sigma_2$ with a block diagonal metric
\beq
g_{\mu\nu} = \left( \begin{array}{cc} g^{(1)} & 0 \\ 0& g^{(2)} \end{array}
\right) ,
\label{block}
\eeq
we get for the effective action for the external gravitational field
\beq
S_{eff}= ({\rm dim} R - {\rm dim} G)
\left(\frac{\chi_1}{48\pi^2} \int_{\Sigma_2}
R^{(2)}\frac{1}{\Delta^{(2)}} R^{(2)} +
\frac{\chi_2}{48\pi^2} \int_{\Sigma_1}
R^{(1)}\frac{1}{\Delta^{(1)}} R^{(1)} \right) ,
\label{effGRr}
\eeq
where $\chi_{1,2}$ are the Euler characteristics of $\Sigma_{1,2} ,$
$R^{(1)},$ $R^{(2)}$ are the Riemann tensors of $\Sigma_{1,2} ,$ and
$\Delta^{(1)} ,$ $\Delta^{(2)}$ are the corresponding Laplace \op s.
We observe that this is a sum of two induced Liouville actions for the
conformal theories on $\Sigma_2$ and $\Sigma_1$ with the central charge
$\c_1$ and $\c_2$ respectively.
If the gauge interactions are absent the factor $({\rm dim} R - {\rm dim}
G)$ changes into $N_f$ (the number of chiral supermultiplets).

Thus the anomaly in the decoupling of the external metric in the heterotic
topological theory can be interpreted as a conformal anomaly in the
2D conformal theories embedded into the manifold $M=\Sigma_1 \times
\Sigma_2 .$
In the
following sections we will demonstrate that such conformal theories
indeed appear in the cohomology of the BRST \op .

\section{Realization of the $W_{1+\infty}$ algebra
in terms of free chiral supermultiplet}
\setcounter{equation}{0}

\subsection{Ghost number $G=0$ cohomology}

In this section we consider a single chiral supermultiplet and
focus to the case of four dimensional manifold
$M=\Sigma_1 \times \Sigma_2 ,$ where $\Sigma_{1,2}$ are 2D Riemann
surfaces.
For simplicity we take $\Sigma_1=T^2$ to be a torus with a flat
two-dimensional metric.
The metric of $M=\Sigma_1 \times \Sigma_2$ can be chosen block diagonal
$g= (g^{(1)},g^{(2)}) ,$ where $g^{(1,2)}$ are two dimensional metrics
for $\Sigma_{1,2}$ respectively.

We introduce below some bosonic $Q$-exact \op s with the ghost number $G=0,$
which are relevant for our construction of the $W_{1+\infty}$ algebra.

The equations of motion for a free chiral supermultiplet read (see
\eq{freeCH})
\beq
D_{\bm} \p_{\bn} -D_{\bn} \p_{\bm} =0,\;\;\;
D^{\bm} \p_{\bm} =0,\;\;\; D_{\bm} \bp^{\bm\bn} + D^{\bn} \bp = 0,
\label{freeEQ}
\eeq
$$D^{\bm}D_{\bm} \h = 0,\;\;\; D^{\bm}D_{\bm} \bar{\h} = 0 .$$
Let us define the following \op s
\beq
W_{n+1} (z,\bar{z}) = 2\pi \sg2 d^2 u ( -\partial_1 \bar{\phi}
\partial^n_1 \phi + g^{2\bar{2}}\bar{\psi}_{12} \partial^n_1
\psi_{\bar{2}} ),
\eeq
where $z$ and $u$ are the complex coordinates on $\Sigma_1$ and
$\Sigma_2$ respectively; $\partial_1 =\partial/\partial z ,$
$\partial_{\bar{1}} =\partial/\partial \bar{z} ,$
$\partial_2 =\partial/\partial u$ and $\partial_{\bar{2}} =
\partial/\partial {\bar{u}} .$
By using the definition of $Q$ in \eq{freeQ}
it is easy to see that these \op s are $Q$-closed on mass-shell
\beq
\{ Q, W_{n+1} \} = 2\pi \sg2 d^2 u ( \partial_1
\bar{\psi} \partial^n_1 \phi - g^{2\bar{2}} \bar{\psi}_{12} \partial^n_1
\partial_{\bar{2}} \phi)=
\eeq
$$=2\pi \sg2 d^4 u (- (\partial_1^n \phi)
(\partial_1 \bar{\psi} + g^{2\bar{2}} \partial_{\bar{2}}
\bar{\psi}_{21})) =0 .$$
Moreover these \op s are holomorphic on $\Sigma_1$ in the $Q$-cohomology
\beq
\partial_{\bar{1}} W_{n+1} = 2\pi \sg2 d^2 u(-\partial_1\partial_{\bar{1}}
\bar{\phi} \partial^n_1 \phi - \partial_1 \bar{\phi}
\partial^n_1\partial_{\bar{1}} \phi + g^{2\bar{2}} (\partial_{\bar{1}}
\bar{\psi}_{12}) \partial^n_1 \psi_{\bar{2}} + g^{2\bar{2}}
\bar{\psi}_{12} \partial^n_1 \partial_{\bar{1}} \psi_{\bar{2}})=
\eeq
$$=2\pi \{ Q, \sg2 d^2 u ( -\partial_1 \bar{\phi} \partial_1^n \psi_{\bar{1}}
- g^{2\bar{2}} \partial_2 \bar{\phi} \partial^n_1 \psi_{\bar{2}} ) \} .$$
Thus we see that $\partial_{\bar{1}} W_{n+1}$ is $Q$-exact and hence
trivial in cohomology.

It is worth emphasizing that the \op s $T=W_2$ and $J=W_1$ are 11- and 1-
components of 2D tensors on $\Sigma_1 .$
These are actually the components of the four dimensional
energy momentum tensor and $U(1)$ current integrated over
$\Sigma_2$ respectively.
This fact is responsible for the holomorphicity of $T$ and $J.$
Indeed the four dimensional current $J_{\mu}$
corresponds to an unbroken $U(1)$ $R$-symmetry of the action and obeys
\beq
\partial_n J^n +\partial^n J_n =0,
\eeq
while for the energy-momentum tensor $\theta_{\mu\nu}$ we have
\beq
\partial^n \theta_{n1} + \partial^{\bar{n}} \theta_{\bar{n} 1} =0 .
\eeq
When integrated over $\Sigma_2$ the terms which are the total
derivatives in $z^2$ and $\bar{z}^{\bar{2}}$ in these
equations vanish.
In turn it is easy to check that the components $J_{\bar{1}}$ and
$\theta_{1\bar{1}}$ are $Q$-exact because of the $Q$-exactness of the
action (for example, $\theta_{1\bar{1}}$ is a variation of the action
with respect to the component $g_{1\bar{1}}$ of the metric).

Similarly we can also define the $Q$-closed \op s which are
local with respect to the coordinates of the Riemann surface
$\Sigma_2$ and are represented by integrals over $\Sigma_1 .$

These \op s do not of course exhaust the full ghost number $G=0$
cohomology of the BRST \op \ $Q.$
In particular there exist also fermionic $Q$-closed
\op s which are represented by
integrals of composite \op s over $\Sigma_2$ ($\Sigma_2$).
We postpone a detailed description of the $Q$-cohomology for a next
publication.

\subsection{Central charge}

We will now show that the \op s $W_{n+1}$ generate the
$W_{1+\infty}$ algebra in the cohomology of $Q.$
To this end let us consider the operator product expansion (OPE) of the \op s
$W_{n+1} .$
We have
\beq
W_{n+1} (z,\bar{z}) W_{p+1} (w,\bar{w}) = (2\pi)^2
\sg2 d^2 u \sg2 d^2 v
\label{OPEfree}
\eeq
$$\left( <\partial^n_1 \phi (z,\bar{z}, u,\bar{u})
\partial_1 \bar{\phi} (w,\bar{w} ,v,\bar{v})>
<\partial^p_1 \phi (w,\bar{w}, v,\bar{v})
\partial_1 \bar{\phi} (z,\bar{z} ,u,\bar{u})> -\right.$$
$$\left. -<\partial_1^n \psi_{\bar{2}} (z,\bar{z} ,u,\bar{u}) \bar{\psi}_{12}
(w,\bar{w} , v,\bar{v})>
<\partial_1^p \psi_{\bar{2}} (w,\bar{w} ,v,\bar{v}) \bar{\psi}_{12}
(z,\bar{z} , u,\bar{u})>\right) +$$
$$+ (2\pi)^2 \sg2 d^2 u \sg2 d^2 v \left( \partial^n_1 \phi (z,\bar{z} ,
u,\bar{u}) \partial_1 \bar{\phi} (w,\bar{w} ,v,\bar{v}) \right.$$
$$\left. <\partial^p_1 \phi (w,\bar{w}, v,\bar{v})
\partial_1 \bar{\phi} (z,\bar{z} ,u,\bar{u})>+\right.$$
$$\left. + \partial^p_1 \phi (w,\bar{w} ,
v,\bar{v}) \partial_1 \bar{\phi} (z,\bar{z} ,u,\bar{u})
<\partial^n_1 \phi (z,\bar{z}, u,\bar{u})
\partial_1 \bar{\phi} (w,\bar{w} ,v,\bar{v})>- \right.$$
$$\left. -\partial_1^n \psi_{\bar{2}} (z,\bar{z} ,u,\bar{u}) \bar{\psi}_{12}
(w,\bar{w} , v,\bar{v})
<\partial_1^p \psi_{\bar{2}} (w,\bar{w} ,v,\bar{v}) \bar{\psi}_{12}
(z,\bar{z} , u,\bar{u})> -\right.$$
$$\left. -\partial_1^p \psi_{\bar{2}} (w,\bar{w} ,v,\bar{v}) \bar{\psi}_{12}
(z,\bar{z} , u,\bar{u})
<\partial_1^n \psi_{\bar{2}} (z,\bar{z} ,u,\bar{u}) \bar{\psi}_{12}
(w,\bar{w} , v,\bar{v})>\right) +$$
$$+ {\rm terms \;\; non-singular \;\;at \;} (z\to w).$$
Here $z$ ($\bar{z}$) and $w$ ($\bar{w}$) are the complex coordinates
on $\Sigma_1$ while $u$ ($\bar{u}$) and $v$ ($\bar{v}$) are the complex
coordinates on $\Sigma_2$ ($\partial_1 =\partial/\partial z$ or
$\partial_1 =\partial/\partial w$ and $\partial_2 = \partial /\partial
u$ or $\partial_2 = \partial /\partial v$).

The propagators of free matter fields are given in the Appendix.
Let us calculate first the most singular (at $z\to w$)
terms proportional to the unity \op \ in \eq{OPEfree}.
These contributions can be represented as follows
\beq
<W_{n+1} (z,\bar{z}) W_{p+1}(w,\bar{w})> =-4\pi^2 \int_{\Sigma_2} d^2 u
\int_{\Sigma_2} d^2 v
\eeq
$$\left[ \left(
\frac{\partial_1^{n+1}}{\Delta_{00}}(z,\bar{z} ,u, \bar{u})
\delta^2 (z-w) \delta^2 (u-v)\right)
\left( \frac{\partial_1^{p+1}}{\Delta_{00}} (w,\bar{w} ,v,\bar{v})
\delta^2 (w-z) \delta^2 (v-u)\right) - \right.$$
$$\left. -\left(
\frac{\partial_1^{n+1}}{\Delta_{01}}(z,\bar{z} ,u, \bar{u})
\delta^2 (z-w) \delta^2 (u-v)\right)
\left( \frac{\partial_1^{p+1}}{\Delta_{01}} (w,\bar{w} ,v,\bar{v})
\delta^2 (w-z) \delta^2 (v-u)\right) \right] ,$$
where
\beq
\Delta_{00} = \partial_1 \partial_{\bar{1}} + e^{-\rho}
\partial_2 \partial_{\bar{2}} ,\;\;\;
\Delta_{01} = \partial_1 \partial_{\bar{1}} + \partial_{\bar{2}}
e^{-\rho} \partial_2
\label{centrW}
\eeq
are the Laplace \op s on $(0,0)$ and $(0,1)$ forms on $M,$
$\exp \rho = g_{2\bar{2}}$ (the metric on $\Sigma_1$ is assumed to be
flat, and $g_{2\bar{2}}$ does not depend on coordinates on $\Sigma_1$).

In order
to compute this contribution we consider an expansion in derivatives
of the external metric in \eq{centrW}
(we can do it in the limit $|z-w|\to 0$).
It is easy to see that for a constant metric ($\rho =const$)
this expression vanishes due to the cancellation between bosonic and
fermionic contributions.
In the linear approximation in $\partial_{\mu} \rho$ we have
\beq
<W_{n+1} (z,\bar{z}) W_{p+1}(w,\bar{w})> =-4\pi^2
\int_{\Sigma_2} d^2 u \int_{\Sigma_2} d^2 v
\label{centrWexp}
\eeq
$$\left[ \left(
\frac{\partial_1^{n+1}}{\Delta_{00}} (\partial_{\bar{2}}\rho)
\partial^{\bar{2}} \frac{1}{\Delta_{00}}(z,\bar{z} ,u, \bar{u})
\delta^2 (z-w) \delta^2 (u-v)\right) \times \right.$$
$$\times \left( \frac{\partial_1^{p+1}}{\Delta_{00}} (w,\bar{w} ,v,\bar{v})
\delta^2 (w-z) \delta^2 (v-u)\right) + $$
$$+\left(
\frac{\partial_1^{n+1}}{\Delta_{00}} (\partial_{\bar{2}}\rho)
\partial^{\bar{2}} \frac{1}{\Delta_{00}}(z,\bar{z} ,u, \bar{u})
\delta^2 (z-w) \delta^2 (u-v)\right) \times$$
$$\left. \times
\left( \frac{\partial_1^{p+1}}{\Delta_{00}} (\partial_{\bar{2}} \rho)
\partial^{\bar{2}} \frac{1}{\Delta_{00}} (w,\bar{w} ,v,\bar{v})
\delta^2 (w-z) \delta^2 (v-u)\right) \right] .$$
We use the fact that
\beq
\frac{1}{\partial_1\partial_{\bar{1}} + \partial_2\partial_{\bar{2}}}
\delta^2 (z) \delta^2 (u) = \frac{1}{4\pi^2} \frac{1}{|z|^2 + |u|^2} ,
\eeq
and integrate over $u-v$ in \eq{centrWexp}.
We get
\beq
<W_{n+1} (z,\bar{z}) W_{p+1}(w,\bar{w})> =
\frac{n!p! (-1)^n}{(z-w)^{n+p+2}}\cdot
\frac{1}{4\pi} \int_{\Sigma_2} d^2 u (\partial_2\partial_{\bar{2}}\rho)=
\label{Wcharge}
\eeq
$$= \frac{n!p!(-1)^{n+1} \chi}{2(z-w)^{n+p+2}} ,$$
where $\chi = (-1/2\pi) \sg2 \partial_2 \partial_{\bar{2}} \rho =2(1-g)$
is the Euler characteristic of the Riemann surface $\Sigma_2 .$

Actually we can do even better and show that this is an exact result which
does not depend on our assumption that the metric changes slowly as
compared to the distance $|z-w| .$
To this end we split the contributions of zero and
non-zero modes of the 2D Laplace \op s
$\Delta^{(2)}_{00} =\exp (-\rho)\partial_{\bar{2}} \partial_2$ and
$\Delta^{(2)}_{01}= \partial_{\bar{2}} \exp (-\rho) \partial_2$
which act on the $(0,0)$ and $(0,1)$ forms on $\Sigma_2$ respectively.
The only zero mode of $\Delta^{(2)}_{00}$ on a compact Riemann surface
$\Sigma_2$ corresponds to a constant wave function $|0>$,
while there are $g$ zero modes of $\Delta^{(2)}_{01}$ with
anti-holomorphic wave functions
$|1_i>,$ $i=1,...,g,$ where $g = {\rm dim} \; H^{1,0}(\Sigma_2).$
Thus we have
\beq
<W_{n+1} (z,\bar{z}) W_{p+1} (w,\bar{w})> = 4\pi^2
\int_{\Sigma_2} d^2 u
\eeq
$$ \left[ <0| \left( \frac{\partial_1^{n+1}}{\lsc} \right)_{z,w}
\left( \frac{\partial_1^{p+1}}{\lsc }\right)_{w,z} |0> - \right.$$
$$\left.
-\sum_{i=1}^{g} <1_i| \left( \frac{\partial_1^{n+1}}{\lvc }\right)_{z,w}
\left( \frac{\partial_1^{p+1}}{\lvc }\right)_{w,z} |1_i> \right] +$$
$$+ 4\pi^2 \lim_{\epsilon \to 0} \int_{\Sigma_2} d^2 u$$
$$\left[ \left( \frac{\sc }{\sc - \epsilon^2}\cdot
\frac{\partial_1^{n+1}}{\lsc } \right)_{z,w}
\left( \frac{\sc }{\sc - \epsilon^2} \cdot
\frac{\partial_1^{p+1}}{\lsc } \right)_{w,z}
\delta^2 (u-v)|_{u=v} - \right. $$
$$\left. - \left( \frac{\vc }{\vc -\epsilon^2}\cdot
\frac{\partial_1^{n+1}}{\lvc } \right)_{z,w}
\left( \frac{\vc }{\vc -\epsilon^2} \cdot
\frac{\partial_1^{p+1}}{\lvc } \right)_{w,z} \delta^2
(u-v)|_{u=v} \right] ,$$
where $(...)_{z,w}$ stands for an \op \ on $\Sigma_2$ which is the
$(z,w)$ matrix element with respect to the coordinates on $\Sigma_1 ,$
i.e., for example,
\beq
\left( \frac{\partial_1^{n+1}}{\lsc} \right)_{z,w} =
\left( \frac{\partial_1^{n+1}}{\lsc} \right) (z,\bar{z} , u, \bar{u})
\delta^2 (z-w)
\eeq
and
\beq
P_{00} = \lim_{\epsilon\to 0} \frac{\sc}{\sc - \epsilon^2} ,\;\;\;
P_{01} = \lim_{\epsilon\to 0} \frac{\vc}{\vc - \epsilon^2}
\eeq
are the projectors to non-zero modes of the \op s
$\sc$ and $\vc$ respectively; $\epsilon$ is a real parameter.

We integrate by parts (on $\Sigma_2$) the second term and observe
that the contribution from the non-zero modes vanishes.
In turn for the contribution from the zero modes we easily get
\beq
4\pi^2 (1-g) \left(
\frac{\partial^{n+1}_1}{\partial_1\partial_{\bar{1}}} \right)
(z,\bar{z}) \delta^2 (z-w) \left(
\frac{\partial^{p+1}_1}{\partial_1\partial_{\bar{1}}} \right)
(w,\bar{w}) \delta^2 (w-z) =
\eeq
$$= (1-g) (-1)^{p+1} (\partial_1^{n+1} \log (z-w))
(\partial_1^{p+1} \log (z-w)) =
\frac{\chi n! p! (-1)^{n+1}}{2(z-w)^{n+p+2}} .$$
We see that this result is exact and agrees with \eq{Wcharge}.
It is worth emphasizing that
this contribution is remarkably holomorphic on $\Sigma_1 .$

\subsection{OPE}

Other terms in the OPE for $W_{n+1} W_{p+1}$ can be represented after
some calculations in the following form (at this point
for simplicity we do not take into account the presence of the external
metric)
\beq
\int_{\Sigma_2} d^2 u \int_{\Sigma_2}
d^2 v \left[ \left( \partial_z^n \phi (z,\bar{z} ,u,\bar{u})
\partial_w \bar{\phi} (w,\bar{w} ,v,\bar{v}) + \right. \right.
\eeq
$$\left. + \partial_z^n
\psi_{\bar{2}} (z,\bar{z} ,u,\bar{u}) \bar{\psi}_{12} (w,\bar{w}
,v,\bar{v}) \right)
\frac{(p+1)! (-1) (\bar{z}-\bar{w})^{p+1}}{(|z-w|^2 +|u-v|^2)^{p+2}}
-$$
$$- \left( \partial_w^p \phi (w,\bar{w} ,v,\bar{v})
\partial_z \bar{\phi} (z,\bar{z} ,u,\bar{u}) + \partial_w^p
\psi_{\bar{2}} (w,\bar{w} ,v,\bar{v}) \bar{\psi}_{12} (z,\bar{z}
,u,\bar{u}) \right)$$
$$\left. \frac{(n+1)! (-1)^{n+1} (\bar{z}-\bar{w})^{n+1}}{(|z-w|^2
+|u-v|^2)^{n+2}} \right] ,$$
where $\partial_z =\partial/\partial z$ and
$\partial_w =\partial/\partial w .$
By expanding the integrand in powers of $z-w$ ($\bar{z}-\bar{w}$) and
$u-v$ ($\bar{u}-\bar{v}$) and integrating over
$u-v$ and $\bar{u}-\bar{v}$ we get
\beq
2\pi \sum^{\infty}_{k=0} \frac{(z-w)^k}{k!} \sg2 d^2 v \left[
\left( \partial_w \bar{\phi} \partial_w^{n+k} \phi - g^{2\bar{2}}
\bar{\psi}_{12}
\partial_w^{n+k} \psi_{\bar{2}} \right) \frac{p! (-1)}{(z-w)^{p+1}}
+\right.
\eeq
$$\left. \left( \partial_w^{k+1} \bar{\phi} \partial_w^p \phi - g^{2\bar{2}}
\partial_w^k \bar{\psi}_{12}
\partial_w^p \psi_{\bar{2}} \right) \frac{n! (-1)^n}{(z-w)^{n+1}}
\right] +$$
$$+ \; Q{\rm -exact \;\; terms}=$$
$$= \frac{p+n}{(z-w)^2} W_{p+n} (w,\bar{w}) + \frac{n}{z-w} \partial_w
W_{n+p} + f(W_{n+p-1} ,..., W_1) +$$
$$+{\rm regular \;\; terms} + \; Q{\rm -exact \;\; terms} .$$
Here we used that $\partial_{\bar{n}} \phi = \{ Q, \psi_{\bar{n}} \} ,$
$\bar{\psi} =\{ Q,\bar{\phi} \}$ and the equations of motion.
Thus the OPE for $W_{n+1} W_{p+1}$ is holomorphic on $\Sigma_1$ in
$Q$-cohomology.

It is easy to see that the \op \ $T=W_2$ generates the
holomorphic Virasoro algebra
with the central charge $\chi$ which is the Euler characteristic of
$\Sigma_2 .$
Moreover one can see that the \op s $W_{n+1}$ generates
the holomorphic $W_{1+\infty}$ algebra.
Indeed by introducing the Fourier modes
\beq
W^n_s = \oint \frac{dz}{2\pi i} z^{n+s} W_{n+1}(z,\bar{z})
\eeq
we get (in the $Q$-cohomology)
\beq
[W^n_s ,W^p_{s'}] = \frac{\chi (-1)^{n+1} n! p!}{2} \cdot
\frac{(s+n)...(s-p)}{(n+p+1)!} \delta_{s+s' ,0} +
(sp -ns') W^{n+p-1}_{s+s'} + R .
\eeq
Here $R$ stands for terms which depend only on $W^k_i$ with $k<n+p-1 .$
The standard $W_{1+\infty}$ commutation relations correspond to $R=0$
(\cite{wintro,center}, see also \cite{bakas}) while in our algebra $R\neq 0.$
However by adding to the \op s $W_{n+1}$ appropriate linear combinations
$\sum_{k=0}^{n-1} a_k \partial_z^{n-k} W_{k+1} ,$ where $a_k$ are
constant coefficients, one can recover the standard
commutation relations with $R=0.$

Similarly we can construct the $W_{1+\infty}$ algebra on the Riemann
surface $\Sigma_2 .$
Its central charge is given by the Euler characteristics of $\Sigma_1 .$
Thus we have a direct sum of two $W_{1+\infty}$ algebras embedded into
the four dimensional theory of free chiral supermultiplet.

\subsection{Ghost number $G\neq 0$}

We considered only the cohomology with the ghost number $G=0.$
For $G\neq 0$ in the free theory one can also define the following \op s
\beq
W^{(n)}_{k+1} = 2\pi \sg2 \left[ -\phi^n \partial_1^{k+1} \bar{\phi} +
g^{2\bar{2}}
n\phi^{n-1} (\partial^k_1 \bar{\psi}_{12}) \psi_{\bar{2}} \right] .
\eeq
These \op s are $Q$-closed and holomorphic in the $Q$-cohomology:
\beq
\{ Q,W^{(n)}_{k+1} \} = 0,\;\;\; \partial_{\bar{1}} W^{(n)}_{k+1} =
\{ Q, ...\} .
\eeq
One can calculate the OPE for these \op s.
In particular the most singular ("central") term in the OPE has the
following form
\beq
W^{(n)}_{k+1} (z, \bar{z}) W^{(m)}_{p+1} (w, \bar{w}) =
\frac{mnk!p! (-1)^{p+1}}{(z-w)^{p+k}}
\eeq
$$\sg2 d^2 u \left( |\omega_0|^2 -\sum_{i=1}^g |\omega_i|^2 \right)
(u,\bar{u}) \cdot \phi^{n-1} (z,\bar{z} ,u,\bar{u}) \phi^{m-1}
(w,\bar{w} ,u,\bar{u}) +...,$$
where $\omega_0$ and $\omega_i ,$ $i=1,...,g,$ are the normalized to one
wave functions of the zero modes of the \op s $\sc$ and $\vc$ respectively.
Unfortunately I do not know of any natural interpretation for this
extended algebra.

\section{Virasoro algebra for a self-interacting chiral supermultiplet}
\setcounter{equation}{0}

Let us now consider a manifold $M=\Sigma_1 \times \Sigma_2$ where
the genera of both Riemann surfaces $\Sigma_{1,2}$ are non-zero.
In this case the manifold $M$ has $H^{2,0} (M) \neq 0.$
Let $E_{mn}$ be a non-trivial holomorphic $(2,0)$ form.
For simplicity we shall also assume that $\Sigma_1 =T^2 .$

In this case it is possible to introduce a superpotential into our
model.
Let us consider the model with a quasihomogeneous superpotential
$W(x)= \lambda x^{N+1}/(N+1) ,$ where $\lambda$ is a coupling
constant (from now on we shall suppress $\lambda$ for simplicity).
We shall show that in this case the $W_{1+\infty}$ algebra is
broken down to a Virasoro one.

The equations of motion in this case read
\beq
\bar{N}_{mn} =-E_{mn} \phi^N ,\;\;\; N_{\bar{m}\bar{n}} = 2
S_{\bar{m}\bar{n}} \bar{\phi}^N ,
\label{freeEq}
\eeq
$$D_{\bar{m}} \bar{\psi}^{\bar{m}\bar{n}} + D^{\bar{n}} \bar{\psi} =
2NE^{\bar{m}\bar{n}}\psi_{\bar{m}} \phi^{N-1} ,\;\;\;
D^{\bar{m}} \psi_{\bar{m}} = -2 S^{12} \bar{\psi}_{12}
N\bar{\phi}^{N-1} ,$$
$$D_{\bar{m}} \psi_{\bar{n}} -D_{\bar{n}} \psi_{\bar{m}}=
2S_{\bar{m}\bar{n}} \bar{\psi} N \bar{\phi}^{N-1} ,$$
$$ D^{\bar{m}} D_{\bar{m}} \phi =-4NS^{12} E_{12}\phi^N \bar{\phi}^{N-1} +
2N(N-1) S^{12} \bar{\psi}_{12} \bar{\psi} \bar{\phi}^{N-2} ,$$
$$ D^m D_m \bar{\phi} =-4NS^{12} E_{12}\bar{\phi}^N \phi^{N-1} -
N(N-1) E^{\bm\bn} \p_{\bm} \p_{\bn} \phi^{N-2} .$$
Here $S_{\bar{m}\bar{n}}$ is a non-singular $(0,2)$ form.

As a consequence of
these equations of motion one can see that the local \op \
$E_{mn} \h^N =0$ in $Q$-cohomology since it is $Q$-exact
\beq
\{ Q, \frac{1}{2} \bp_{mn} \} = -\bar{N}_{mn} = E_{mn} \h^N .
\eeq
This is a four-dimensional analog of the relations that define the ground ring
in 2D $N=2$ supersymmetric theories.
If the (0,2)-form $E_{mn}$ has no zeros (i.e. $\Sigma_1 =\Sigma_2 =
T^2$) we get
\beq
\h^N =0 .
\eeq
Hence the algebra of local scalar \op s is generated by monomials $\h^k
,$ $k =0,...,N-1$ modulo $\h^N .$
In the particular case of $N=1$ which corresponds to the free massive theory
the \op \ $\h =0$ in the $Q$-cohomology.
In turn as we shall see below in this case the central charge of a Virasoro
algebra which appears in the $Q$-cohomology turns out formally to be
zero.
Actually in such a special case the corresponding stress turns out
to be $Q$-exact due to the equations of motion \rf{freeEq}.

In the case when $E_{mn}$ has zeros the analysis of the algebra of local
scalar \op s is not so straightforward because the \op \ $\bp_{mn}/E_{12}$
is not well defined due to a singularity in $1/E_{12} .$
Correspondingly a Virasoro algebra which appears in the $Q$-cohomology
has a non-zero central charge even in the case $N=1.$

We now construct a Virasoro algebra on $\Sigma_2 .$
It is convenient to introduce the following \op s
\beq
W_{p,q}= \sg2 d^2 u \left( - \partial_1^{q+1} \bar{\phi} \partial^p_1 \phi +
g^{2\bar{2}}
\partial^q_1 \bar{\psi}_{12} \partial^p_1 \psi_{\bar{2}} \right) .
\eeq
By using the equations of motion (4.1) and taking $E_{12}$ to be independent
of the coordinates on $\Sigma_1$ one can check that
\beq
\{ Q, W_{pq} \} = 2 \sg2 d^2 u \; E_{12} g^{2\bar{2}}
\left[ N (\partial^p_1 \phi) \partial_1^q
(\psi_{\bar{2}} \phi^{N-1}) + (\partial^q_1 \phi^N) (\partial^p_1
\psi_{\bar{2}})\right] .
\eeq
This expression is $Q$-closed but non-vanishing, and hence
the \op s $W_{p,q}$ do not belong to the $Q$-cohomology.

Let us consider a linear combination of $W_{p,q}$ with a definite
spin $n$
\beq
\tilde{W}^{(n)} = \sum_{p+q=n} a^{(n)}_{p,q} W_{p,q} ,
\eeq
where $a^{(n)}_{p,q}$ are some constants.
We want to find out all the sets of $a^{(n)}_{p,q}$ for which
$\{Q,\tilde{W}^{(n)} \} =0 .$
We will actually show that the only non-trivial solution appears for
$n=1.$
To show this let us formally consider the integrals of $\tilde{W}^{(n)}$
over the holomorphic coordinate $z_1$ on $\Sigma_1$ (with vanishing
boundary \cn s)
\beq
A^{(n)}=\int dz_1 \tilde{W}^{(n)} = \sum_{p+q=n} a^{(n)}_{p,q} (-1)^q
\int dz_1 W_{n,0} = b \int dz_1 W_{n+1} .
\eeq
We then have
\beq
\{ Q,A^{(n)}\} = 2b \int dz_1 \sg2 E_{12} g^{2\bar{2}}
[ N (\partial^n_1 \phi) \psi_{\bar{2}}
\phi^{N-1} + \phi^N \partial_1^n \psi_{\bar{2}} ].
\eeq
This expression vanishes only at $b=0$ or at $n=1$ (in the latter case
the integrand is a total derivative in $z_1$).
At $b=0$ however the \op \ $\tilde{W}^{(n)}$ is a total derivative in
$z_1$ since its formal integral over $z_1$ vanishes, and therefore it
reduces to the \op s $\tilde{W}^{(k)}$ with $k<n.$
We thus have inductively shown that the only non-trivial $Q$-closed
\op \ $\tilde{W}$ can appear at $n=1 .$
It is easy to check that for any $N$ such an \op \ does exist and it has
the following form
\beq
T_N=W_2 - \frac{1}{N+1} \partial_1 W_1 ,
\eeq
so that
\beq
\{ Q,T_N \} =0, \;\;\; \partial_{\bar{1}} T_N = \{ Q,...\} .
\eeq
Notice that the 1-component of the $U(1)$ current does not belong
to the $Q$-cohomology since the phase symmetry is broken by the
superpotential.

The \op \ $T_N$ has a spin 2 and generates the holomorphic Virasoro
algebra on $\Sigma_1 .$
In order to check it it is necessary to calculate the OPE for $T_N (z,\bar{z})
T_N (w,\bar{w}) .$
It can be done in the limit of weak coupling similarly with the calculation
of Ref.\cite{WiLG,SiWi}.
The point is that the superpotential terms in the Lagrangian are
dimensionful (for any superpotential in contrast to the untwisted
supersymmetric theory)
and make less singular contributions to the OPE as compared
to the free ones.
Formally the same follows from the fact
that the OPE for this \op \ can not depend on a choice of the form
$S_{\bm\bn}$ in the $Q$ cohomology since the terms in the Lagrangian
which depend on $S_{\bm\bn}$ are $Q$-exact (see \eq{selfint}).
In particular it is easy to calculate the central charge
which is given by
\beq
c_N= \chi \left( 1-\frac{6}{N+1} +\frac{6}{(N+1)^2} \right) .
\label{cVir}
\eeq
The model is renormalizable only if $N=1,2.$
For the case $N=1$ which corresponds to the free massive model
we have $c_2=-\chi/2 = g-1 ,$ while for $N=2$ $c_3 =-\chi/3 =2(g-1)/3 .$
In both cases the central charge is non-negative since we assumed
that $g>0 .$

Let us now discuss the behaviour of this algebra with respect to the
renormalization group.
The superpotential is not $Q$-exact.
Hence the physical correlators can depend on the coupling constant in
the superpotential.
This is a difference of the present case from the twisted $N=1$
gauge theory without a superpotential for the matter fields
\cite{ajoh} where all the action is $Q$-exact (see below for a
discussion).
However the superpotential is not renormalizable
\cite{book} while the $D$-terms which are $Q$-exact are only renormalizable.
We thus conclude that our construction is invariant under the
renormalization group.
This fact can also be understood as follows.
The quantum effects result in only a renormalization of the wave
functions by a factor $Z.$
In turn the same factor $Z$ appears in the \op \ $T_N$ due to quantum
effects.
These effects are non-trivial for the \op \ $T_N$ because it is
$Q$-closed only on mass-shell in contrast to the case of \op s
$\phi^k$ which are $Q$-closed off mass-shell and do not acquire any
multiplicative factor under renormalization \cite{johRG}.
Notice that since the \op \ $T_N$ is an integral of a component of the
conserved energy-momentum tensor it does not acquire any its own
renormalization factor.
After a redefinition of the quantum fields the factor $Z$ disappears
in the OPE for $T_N$ which is therefore invariant under the
renormalization group.

Notice that similarly we can construct the Virasoro algebra on the
Riemann surface $\Sigma_2 .$
The central charge of this algebra will be given by the same expression
\eq{cVir} with a change of the Euler characteristics of the surface
$\Sigma_2$ into that of $\Sigma_1 .$
Thus we have a direct sum of two Virasoro algebras embedded into the
four dimensional theory.

\section{A twisted supersymmetric gauge theory}
\setcounter{equation}{0}

In this section we construct the Virasoro algebra in the $Q$-cohomology
of a heterotic topological gauge theory without a superpotential for the
matter fields
and discuss the instanton effects and a possible relation to two
dimensional supersymmetric sigma models.

\subsection{Gauge fixing and equations of motion}

In order to formulate the theory at the quantum level one has to
fix a gauge.
To this end we introduce the ghost field $c,$ the antighost
field $c^+$ and an auxiliary field $B.$
The BRST transformations responsible for the gauge fixing read
\beq
Q_{BRST} A_n = D_n c ,\;\;\; Q_{BRST} A_{\bar{n}} = D_{\bar{n}} c
,\;\;\; Q_{BRST} \chi_n = i \{ c ,\chi_n \} ,
\eeq
$$Q_{BRST} \bar{\lambda} = i \{ c, \bar{\lambda} \} ,\;\;\;
Q_{BRST} \bar{\lambda}^{mn} = i \{ c, \bar{\lambda}^{mn} \} ,\;\;\;
Q_{BRST} D' = i [c,D'] ,$$
$$Q_{BRST} c = \frac{i}{2} \{ c,c\} ,\;\;\; Q_{BRST} c^+ =B ,\;\;\; Q_{BRST}
B =0.$$
The generator of this symmetry anticommutes with the generator $Q$ of
the twisted supersymmetry
(the generator $Q$ is assumed to annihilate the fields $c,$ $c^+$ and
$B$).
Therefore the total symmetry of the theory at the quantum level should
be generated by $Q' = Q+Q_{BRST} .$
Let us fix the gauge by adding the following $Q+Q_{BRST}$-exact \op \
to the Lagrangian
\beq
L_{\rm fix} = {\rm Tr}
\{ Q' , c^+ \partial^{\mu} A_{\mu} + \frac{1}{2\alpha} c^+ B \} =
\label{Lfix}
\eeq
$$={\rm Tr} [B \partial^{\mu} A_{\mu} - c^+ \partial^{\mu} D_{\mu} c -
c^+ \partial^m \chi_m + \frac{1}{2\alpha} B^2 ],$$
where $\alpha$ is a gauge fixing parameter.
The term $c^+ \partial^m \chi_m$ is unusual for gauge fixing.
To recover the background field gauge one has to change the usual
derivatives $\partial_{\mu}$ in the above equations into $D^{(0)}_{\mu}=
\partial_{\mu} -i A^{(0)}_{\mu} ,$ where $A^{(0)}_{\mu}$ is an external
gauge field.
For definiteness we shall fix the gauge parameter $\alpha= -1.$

For the total Lagrangian $L_{tot} = L + L_{{\rm fix}} ,$ where $L$ is
given by \eq{totalL} the equations of motion for the gauge
supermultiplet read
\beq
(2D_m F^{mn} + iD^n D'  - \{ \bl^{mn} ,\c_m \} + D^{(0)n} B - i \{
D^{(0)n} c^+ ,c\})^a +
\label{eqmotion}
\eeq
$$+ i \bp t^a \p^n +i\bh t^a D^n \h =0,$$
$$(2D^m F_{mn} - iD_n D'  + \{ \bl ,\c_n \} + D^{(0)}_n B - i \{
D^{(0)}_n c^+ ,c\})^a + $$
$$+i \bp_{nm} t^a \p^m -iD_n \bh t^a \h =0 , $$
$$D'^a = ig^{\bm m} F^a_{\bm m} + i \bh t^a \h ,\;\;\;
iD_m \c_n - iD_n \c_m - \frac{1}{2} \bp_{mn} t^a \h =0,$$
$$iD^m\c_m +i \bp \h =0,\;\;\; iD_m \bl^{mn} +iD^n \bl -i \bh \p^n = 0,$$
where $t^a$ stands for a generator of the gauge group, while for the
multiplet of matter we have
\beq
D^{\bm} D_{\bm} \h - i \chi_n \p^n + i D' \h = 0, \;\;\;
D_{\bm} D^{\bm} \bh + \frac{1}{4} \bp_{mn} \bar{\lambda}^{mn} +
i\bh D' - i\bp \bl =0 ,
\label{mattermotion}
\eeq
$$D_{\bm} \p_{\bn} -D_{\bn} \p_{\bm} + \frac{1}{2} \bl_{\bm\bn} \h =0,\;\;\;
D^{\bm} \p_{\bm} -i \bl \h =0,\;\;\;
D_{\bm} \bp^{\bm\bn} + D^{\bn} \bp + i \bh \c^{\bn} =0.$$
Thus the quantized theory is invariant under a combined BRST
transformations.

\subsection{Physical correlators}

It follows from section 2. that the physical correlators are sections of a
holomorphic bundle on $M.$

Let us consider the structure of a physical correlator.
We observe that the term $(1/2)TrD'^2$ in the Lagrangian \rf{totalL}
is BRST exact and can be taken out from the
Lagrangian without any change of the correlator.
After this modification one can integrate out over the field $D'$
that leads to the following constraint
\beq
g^{m\bar{n}} F_{m\bar{n}} =0.
\label{firstcon}
\eeq
It is easy to see that this \cn \ is necessary for the anti-self-duality
of the Yang-Mills field.
In order to show this it is convenient to use the identity
$\epsilon_{m \bar{n} k \bar{l}} =
g_{m \bar{l}} g_{k \bar{n}} -g_{m \bar{n}} g_{ k \bar{l}} $.
One can then check that
\beq
\epsilon_{\bar{n} km \bar{l}} F^{m \bar{l}}
= - F_{\bar{n} k} - g_{\bar{n} k} (g^{\bar{p} q} F_{\bar{p} q}),
\eeq
$$\epsilon_{\bar{n} km \bar{l}} F^{km} = 2 F_{\bar{n}\bar{l}},
\;\;\; \epsilon_{\bar{n} km \bar{l}} F^{\bar{n}\bar{l}} = 2 F_{km}.$$
In turn the anti-self-duality \cn \ means that
\beq
\epsilon_{\bar{n} km \bar{l}} F^{m \bar{l}} = - F_{\bar{n} k} ,\;\;\;
\epsilon_{\bar{n} km \bar{l}} F^{km} = -2 F_{\bar{n}\bar{l}},\;\;\;
\epsilon_{\bar{n} km \bar{l}} F^{\bar{n}\bar{l}} = -2 F_{km}.
\label{selfcon}
\eeq
{}From these equations it is easy to see that the anti-self-duality \cn \
is equivalent to the following equations:
\beq
g^{\bar{p} q} F_{\bar{p} q} =0,\;\; F^{km} = F^{\bar{n}\bar{l}} =0.
\eeq
\eq{firstcon} coincides with the first of the anti-self-duality \cn s.
The other \cn s in eqs.~\rf{selfcon}
appear in the limit of a weak gauge coupling constant
because in this case the functional integral is dominated by
the fields with $F^{km} = F^{\bar{n}\bar{l}} =0$
which correspond to the minimum of the action.
In turn we can consider this limit since the action of the theory is
BRST exact and, hence,
the correlators are independent of the value of the coupling constant.
The theory is therefore similar to the topological Yang-Mills theory
\cite{witten} and
the physical correlators can be computed
semiclassically in the presence of an anti-instanton field.

In such a case some of the fields of the model have
zero modes.
These zero modes are absorbed by the \op s inserted into the correlator.
Technically the wave functions of the zero modes
should be substituted into the preexponential factor
in the path integral for the correlator (directly or by using Yukawa
couplings).
Moreover one should integrate over quadratic fluctuations
near the anti-instanton field.
Notice that the classical action \rf{totalL}
for the anti-instanton equals zero.

We assume that the anti-instanton configuration is non-trivial
only for the gauge field while the other fields vanish classically.
Let us analyse the \zm s (for more details see ref.\cite{ajoh}).
Actually it is easy to see from eqs.~\rf{selfcon} that the variation of the
self-duality equations in the vector field $A_n$
(fixing the gauge of the variation
of the gauge field by a \cn \ $D^{\mu} \delta A_{\mu} =0$) gives the following
equations
\beq
D_{[m } \delta A_{n]} = 0,
\;\; g^{\bar{m} n} D_{\bar{m}} \delta A_{n} = 0 ,
\label{eqgluon1}
\eeq
\beq
D_{\bar{[m}} \delta A_{\bar{n]}} =0,\;\;\;
g^{\bar{m} m}D_m \delta A_{\bar{m}} .
\label{eqgluon2}
\eeq
These equations determine the zero modes
of the gauge field near the anti-instanton field up to gauge
transformations.

The equations of motion for the field $\chi_n$
coupled to the anti-instanton field read as follows
\beq
D_{[m } \chi_{n]} = 0,\;\;
g^{\bar{m} n} D_{\bar{m}}\chi_{n} = 0.
\label{eqgluino}
\eeq
Due to the similarity of these equations to eqs.~\rf{eqgluon1}
and \rf{eqgluon2} we see that the
\zm s of the fermionic field $\chi_n$ and $A_n$ coincide.
Therefore the \zm s of $\chi_n$ correspond to a half of tangent vectors to the
moduli space ${\cal M}$ of the anti-instanton because there are no
superpartners
to $A_{\bar{n}} .$

The zero modes of the gauge field near the anti-instanton configuration
determine the collective coordinates which correspond to coordinates in
the moduli anti-instanton space.
For a compact manifold $M$
the moduli space ${\cal M}$ of the anti-instanton is a manifold of dimension
\cite{ahsinger,freed}
\beq
d= p_1(G) -\frac{1}{2} {\rm dim} G (\chi +\tau) ,
\eeq
where $p_1(G)$ is the first Pontryagin class of the adjoint bundle over $M,$
$G$ stands for the gauge group, $\chi$ is the Euler characteristic of
$M,$ and $\tau$ is the signature of $M$
\beq
\chi = \frac{1}{128\pi^2} \int_M \epsilon^{\mu\nu\alpha\beta}
\epsilon^{\gamma\delta\alpha\beta} R_{\mu\nu\alpha\beta}
R_{\gamma\delta\alpha\beta} ,\;\;\;
\tau= \frac{1}{96\pi^2}\int_M \epsilon^{\mu\nu\alpha\beta}
R_{\mu\nu\lambda\rho} R^{\lambda\rho}_{\alpha\beta} .
\eeq
For the case of a K\"ahler manifold $M$ the moduli space ${\cal M}$
is a complex manifold of a complex dimension $d/2$
\cite{itoh,harvey}.
The K\"ahler structure on ${\cal M}$ is induced by a K\"ahler structure
on $M.$
The corresponding K\"ahler form on ${\cal M}$ turns depend on
only the K\"ahler class of the K\"ahler metric on $M$ (see also
\cite{ajoh}).

In this paper we focus to the case of a
generically irreducible anti-instanton field
and assume that only vector fields can have zero modes.

Notice that the fermionic field $\chi_m$ has $d/2$ zero modes
but not $d$ despite the similarity of eqs.~\rf{eqgluon1},\rf{eqgluon2}
and \rf{eqgluino}.
The point is that for the gluonic field we should consider the pairs
$(\delta A_m ,\delta \bar{A}_{\bar{m}})$
while the fermionic field has only components with a holomorphic index.
It is easy to show that if $(\delta A_m ,\delta \bar{A}_{\bar{m}})$
is a wave function for a gluonic zero mode then so is
$(i\delta A_m ,-i\delta \bar{A}_{\bar{m}})$ as it
can be rewritten as $J^{\mu}_{\nu} \delta A_{\mu}$ ($J^{\mu}_{\nu}$ is
the complex structure)
which satisfies the same equation with $\delta A_{\mu}$ as the tensor
$J^{\mu}_{\nu}$ is covariantly constant.

Let us now consider the matter sector of the fields.
The equations of motion for the fermionic vector field $\psi_{\bar{m}}$
which is coupled to the anti-instanton field read
\beq
D_{\bar{m}} \psi_{\bar{n}} -D_{\bar{m}} \psi_{\bar{n}} =0, \;\;\;
D^{\bar{m}} \psi_{\bar{m}} =0 .
\eeq
These equations are similar to those for the zero modes of the gauge
field but
the $\p_{\bn}$ matter field can belong to an arbitrary representation $R$ of
the gauge group (not necessarily to an adjoint one).
The number of zero modes of the field $\p_{\bn}$ depends on the representation
$R$ and is determined by the index theorem for a corresponding Dirac
operator \cite{gilkey}.

We also have to consider the integration over quadratic fluctuations around
the anti-instanton field.
The result of such an integration provides us with a combination of
determinants of the Laplace-type \op s.
As a result we get for the physical correlator a representation as
an integral over moduli space ${\cal M}$ of anti-instanton
of a $(d/2,d/2)$ form on ${\cal M}$ \cite{ajoh}.

\subsection{$U(1)$ and Konishi anomalies}

We now want to study the cohomology of the BRST \op .
We will not give here an exhaustive analysis (it will be given
elsewhere).
Instead we will consider a part of this cohomology for the ghost number
$G=0$ which is relevant for
the construction of the conformal algebra and for an analysis of some
dynamical properties of the theory which is given below.

To this end we consider a twisted formulation of the Konishi anomaly
\cite{konishi}.
Let the matter multiplet transform as a representation $R$ of the gauge
group which consists of irreducible representations $R_i ,$ $i=1,...,f.$
Let a representation $R_i$ enter $R$ with a repetition $n_i .$

Let us first consider the case of massless matter fields.
For a particular irreducible representation $R_i$
at the classical level we have
\beq
\{ Q, \bp^I_{12} \h^I \} =0,
\eeq
where the index $I$ stands for the $I$'s copy of the representation
$R_i$ in $R.$
At the quantum level we have to introduce an ultra-violet
regularization.
To this end it is convenient to use the Pauli-Villars ultraviolet
regularization for the matter multiplet in external gauge
($A^{cl}_{\mu}$) and gauigino ($\c_n^{cl}$) fields.
The Lagrangian for the regulator fields of matter reads
(a subscript $reg$ stands in order to label the regulator fields) read
\beq
L_{reg}^m= \sqrt{g} ( \bar{\phi}_{reg} D^{\bar{m}} D_{\bar{m}}
\phi_{reg} +
\bar{\psi}_{reg}^{\bar{m}\bar{n}} D_{\bar{m}} \psi_{reg,\bar{n}} +
\bar{\psi}_{reg} D^{\bar{m}} \psi_{reg,\bar{m}} + N_{reg,\bar{m}\bar{n}}
\bar{N}^{\bar{m}\bar{n}}_{reg} -
\label{regm}
\eeq
$$-i\bh_{reg} \c_n^{cl} \p^n_{reg} - e^2 \bh_{reg} F^m_{\; m} \h_{reg} )
+$$
$$+\frac{1}{2}ME_{mn} \sum_i
\sum_{IJ} a^{(i)}_{IJ} (\p_{reg}^{I,m}\p_{reg}^{J,n} + N^{I,mn}_{reg}
\h_{reg}^J) -\frac{1}{2}MS_{\bm\bn} \sum_i
(\sum_{IJ} a^{(i)}_{IJ} \bp^{I,\bm\bn}_{reg}
\bp_{reg}^J + 2 \bar{N}^{I,\bm\bn}_{reg} \bh_{reg}^J ) .$$
Here $S_{\bm\bn}$ stands for a non-singular $(0,2)$ form on the manifold
$M;$
$F_{\bm m}$ is the strength of external gauge fields and the covariant
derivatives are assumed to be taken in external gauge fields;
the
indices $I$ and $J$ stand for copies of $R_i$ if the repetition of $R_i$
in $R$ is more than 1.
We assume that the fields which do not have indices $I$ and $J$
(the first two lines in \eq{regm}) correspond to the total
representation $R .$
We also assume a gauge invariant scalar product
of the fields in the mass terms.
$a^{(i)}_{IJ}$ is an appropriate non-degenerate constant matrix defined
above (its choice depends on the representation $R_i$).
For example for the
adjoint representation of the matter field it can be taken to be unity;
if we consider a model with the $SU(2)$ gauge group and the matter
multiplet given by four copies of the spinor representation of $SU(2)$
then $a^{(i)}_{IJ}$ is to be a non-degenerate antisymmetric $4\times 4$ matrix.
We normalize it by the \cn \
$$\sum_J a^{(i)}_{IJ} a^{(i)}_{JK} =\delta_{IK}$$
For simplicity we shall restrict ourselves to the case of the forms
$E_{mn}$ and $S_{\bm\bn}$ being non-zero everywhere (this corresponds to the
case $\Sigma_1=\Sigma_2 =T^2$) and $E_{12} S^{12} = -1.$
The mass terms of the regulator fields
are gauge invariant under simultaneous gauge
transformations of the background and regulator fields.
Moreover they are invariant under the BRST transformations defined in
the same manner as for the physical fields.

We define the regularized \op \
\beq
(\bp^I_{12} \h^I)_{reg} =\bp_{12}^I \h^I -\bp^{reg,I}_{12} \h_{reg,I}.
\eeq
Due to the presence of regulator fields, we now have
\beq
\{ Q, (\bp^I_{12} \h^I)_{reg} \}= 2\bar{N}^{reg,I}_{12} \h_{reg,I} =
ME_{12} \sum_J a^{(i)}_{IJ} \h^I \h^J .
\eeq
By integrating over the regulator fields we get
\beq
\{ Q, (\bp^I_{12} \h^I)_{reg} \}=- \frac{1}{4\pi^2} {\rm Tr}_{R_i}
\c_1\c_2 ,
\label{akon}
\eeq
where ${\rm Tr}_{R_i}$ stands for the trace in the representation $R_i .$
The \op \ ${\rm Tr} \c_1\c_2$ coincides with the \op \
$O^{(0)}_{12}$ defined in eq.(2.24) and is $Q$-closed.
However \eq{akon} shows that it is $Q$-exact.
Hence the physical correlators in the massless theory with insertions of
the \op \ $O^{(0)}_{12}$ vanish.

Let us now consider the theory with massive matter fields.
The mass terms for the physical fields
have the following form (we assume that $H^{2,0} (M) \neq
0;$ see \eq{selfint} with a quadratic superpotential)
\beq
\frac{1}{2}mE_{mn}
\sum_i \sum_{IJ} a^{(i)}_{IJ} (\p^{I,m}\p^{J,n} + N^{I,mn}
\h^J) -\frac{1}{2}mS_{\bm\bn} \sum_i
(\sum_{IJ} a^{(i)}_{IJ} \bp^{I,\bm\bn}
\bp^J + 2 \bar{N}^{I,\bm\bn} \bh^J ) .
\eeq
Here $E_{mn} \in H^{2,0} ,$
$S_{\bm\bn}$ stands for a non-singular $(0,2)$ form on $M;$
$m$ is a mass parameter,
$a_{IJ}$ is an above defined constant matrix;
the indices $I$ and $J$ stand for copies of $R_i$ as long as
the repetition of $R_i$ in $R$ is more than 1.

In the massive theory when the matter fields have non-trivial mass terms
instead of \eq{akon} we get the twisted formulation of the Konishi anomaly
\beq
\frac{m}{2} E_{12} \sum_{J} a^{(i)}_{IJ} \h^I \h^J =
\frac{1}{4\pi^2} {\rm Tr}_{R_i} \c_1\c_2 .
\eeq
In a particular case of matter in the adjoint representation we have
\beq
\frac{m}{2} E_{12} {\rm Tr}_{Ad} \h \h =
\frac{1}{4\pi^2} {\rm Tr}_{Ad} \c_1\c_2 .
\label{kon}
\eeq
Hence in the massive theory the \op \ $O^{(0)}_{12}$ is not $Q$-exact.

The Konishi anomaly has the following consequence.
Let us consider a correlator of $N_{\h}$ local \op s $<\h\h>$
($<...>$ stands for a gauge invariant scalar product in the
representation $R$) and $N_{\c}$
local \op s ${\rm Tr} \; \c_1 \c_2 .$
Such a correlator is a holomorphic form on $M.$
Due to \eq{kon} the tensor structure of the correlator is factorized
out.
As a consequence it reduces to a correlator of $N_{\h}
+N_{\c}$ \op s $<\h\h > .$
In the case of a massless theory the correlator with insertions of
the \op \ $O^{(0)}_{12}$ vanishes due to a triviality of $O^{(0)}_{12}$
in $Q$-cohomology.
On the other hand it vanishes due to the Lorentz invariance of the
theory.
To be more precise one can consider an infinitesimal
holomorphic transformation of
complex coordinates $z_1$ and $z_2$ on the manifold $M.$
It is easy to check that the variation of the action is $Q$-exact under
such a transformation since it reduces to a variation of the external metric.
Hence the correlator does not change.
However the \op s that enter the correlator modify their coordinate
dependence.
Therefore the correlator should be a form on $M$ which is
invariant under such a holomorphic transformation.
Obviously the only non-trivial invariant form is a constant scalar.
We thus maintain the above conclusion of triviality of a correlator with
insertions of $O^{(0)}_{12} .$
In the case of massive matter fields the variation of the action under
holomorphic transformations of coordinates is $Q$-closed but not
$Q$-exact.
Therefore the correlator can be non-vanishing.
The same holds true for the correlators with insertions of the \op \ (2.29)
constructed out of the fields of matter.

Let us now
consider once again a correlator of $N_{\h}$ local \op s $<\h\h >$ and
$N_{\c}$ local \op s $\c_1 \c_2$ in the massless theory.
In the weak coupling limit it can be calculated
by a substitution of the zero modes of the fields $\p_{\bn}$ and $\c_n$ in
an instanton field.
It is tempting to interpret the vanishing of this correlator as a
manifestation of a lack of fields in the preexponential factor which
could absorb these zero modes.
Since the number of zero modes of the fields $\p_{\bn}$ and $\c_n$ is
fixed by the index theorem \cite{gilkey}
one can extract a relation between the Euler characteristic and the
signature of the manifold $M.$
Unfortunately this argument is rather speculative because the vanishing
of the correlator in the ultraviolet limit can be due to an integration
over the moduli space of the anti-instanton.

Let us now restrict ourselves to the case of the four-dimensional manifold
$M= \Sigma_1 \times \Sigma_2 ,$ where $\Sigma_{1,2}$ are two-dimensional
Riemann surfaces.
We shall assume the external metric to be block-diagonal (see \eq{block}).
One can construct $U(1)$ currents
which are in the BRST cohomology at the classical level.
Let us consider the massless theory.
If we take one ($R_i$) of irreducible representations of matter fields
which enter the model one can define for the fields which
belong to the $I$'s copy of the
representation $R_i$
representation the singlet following current
\beq
J^{(I)}= 2\pi \sg2 d^2 u (D_1 \bh^I \cdot \h^I - \bp_{12}^I \p^{I,2} ).
\label{current}
\eeq
At the classical level by using the equations of motion one can check that
$\{ Q,J^{(I)}\} =0$ and $\partial_{\bar{1}} J^{(I)} =\{ Q,...\} .$
However this current contains a chiral fermionic current in terms of untwisted
fields and hence may have an anomaly at the quantum level.
We shall show below that this current is indeed anomalous.

To this end it is convenient to use the Pauli-Villars regularization
introduced above.
We define the regularized current as
\beq
J^{(I)}= 2\pi \sg2 d^2 u (D_1 \bh^I \cdot \h^I - \bp_{12}^I \p^{I,2} ) -
2\pi \sg2 d^2 u (D_1 \bh^I_{reg} \cdot \h^I_{reg} -
\bp^I_{reg,12} \p^{I,2}_{reg}).
\eeq
It is easy to check by using the equations of motion for the Lagrangians
\rf{totalL} and \rf{regm} that
\beq
\{ Q, J^{(I)} \} = 2\pi M
\sg2 d^2 u E_{12} \sum_J a_{IJ} (\h^I_{reg}\p^{J,2}_{reg} +
\p^{I,2}_{reg} \h^J_{reg}) .
\eeq
By using the vertex $-\sum_{I} i\bh_{reg}^I \c_n \p_{reg}^{I,n}$
in the Lagrangian
\rf{regm} one can integrate over the regulator fields and get the
anomaly for the current $J^{(I)}.$
The anomaly is obviously proportional to a Dynkin
index of the representation $R_i .$
Therefore in order to keep our computation as simple as possible
we consider a matter multiplet
in the adjoint representation of the gauge group
(this matter would correspond to the twisted $N=2$
supersymmetric Yang-Mills theory)
while the correct answer for the anomaly of the current constructed out of
the matter fields in a representation $R_i$ is given by the anomaly
for the matter in the adjoint representation times a factor
$T(R_i)/C_2(G) .$

In the case of matter in an adjoint representation the matrix $a_{IJ}$ is
reduced to 1.
In this case ($J^{(I)}\to J^{Ad}$)
\beq
\{ Q, J^{Ad} \} =4\pi M \sg2 d^2 u E_{12} {\rm Tr}_{Ad}
\h_{reg} \p^2_{reg} .
\eeq
By using the propagators of the regulator matter fields
in the external gauge field (the propagators for massive
fields in the external gauge field are given in Appendix) we get
for a current of matter in an irreducible representation $R_i$
\beq
\{ Q, J^{(I)} \}= \frac{T(R_i)}{C_2(G)} \cdot \frac{1}{2\pi}
\int_{\Sigma_2}
d^2 u {\rm Tr}_{Ad} (\c_2 F_{\bar{2} 1} -\c_1 F_{\bar{2} 2}) =
\label{acur}
\eeq
$$=\frac{1}{2\pi} {\rm Tr}_{R_i} \int_{\Sigma_2}
d^2 u (\c_2 F_{\bar{2} 1} -\c_1 F_{\bar{2} 2}).$$
This anomaly is proportional to the \op \ $H^{(1)}_{12,\bar{2}}$ (see
eq.(2.27)) integrated over $\Sigma_2 .$
It is easy to see that this \op \ is $Q$-closed.
Actually \eq{acur} shows that this \op \ is $Q$-exact.
And hence (in the massless theory) the physical correlators with insertions
of this \op \ should vanish.

Let us now consider an anomaly in $\partial_{\bar{1}} J^{(I)}.$
By similar calculations one can show that
\beq
\partial_{1} J^{(I)} = \frac{1}{2\pi} \sg2 d^2 u
H^{(2)}_{12,\bar{1}\bar{2}} .
\eeq
This result agrees with eq.(2.26) (the \op \ $H^{(2)}$ is defined in
eq.(2.27)).
Indeed
\beq
\{ Q, \partial_{1} J^{(I)} \} =\frac{1}{2\pi} \sg2 d^2 u
\{ Q,H^{(2)}_{12,\bar{1}\bar{2}} \} =
\eeq
$$=\frac{1}{2\pi} \sg2 d^2 u \partial_{\bar{1}} H^{(1)}_{12,\bar{2}} =
\partial_{\bar{1}} \{ Q,J^{(I)} \} .$$

\subsection{Holomorphic currents and stress tensor on $\Sigma_2$}

In this section and in the next one we will consider the massless theory
and we focus to the case of four-dimensional manifold
$M= \Sigma_1 \times \Sigma_2 .$

In the case when the matter multiplet does not belong to the adjoint
representation of the gauge group one can still take non-anomalous
linear combinations of the currents $J^{R_i} .$
We can also define non-singlet (with respect to the flavour group
symmetry) currents.
Let the representation of the matter fields be a sum of irreducible
representations $R_i$ with a repetition $n_i.$
It is easy to see that the non-anomalous flavour group symmetry is
\beq
G_F = SU(n_1) \times ... \times SU(n_f) \times U(1)^{n -1} ,
\label{fgroup}
\eeq
where $f$ is a number of different irreducible representations that
enter the model and $n =\sum_i n_i .$
We introduce the multiplet of currents constructed out of the fields
which belong to the representation $R_i$ as follows
\beq
J_{(i)}^{a} = 2\pi \sg2 d^2 u
(T^{(i)})^a_{IJ}(D_1 \bh^I \cdot \h^J - g^{2\bar{2}}\bp_{12}^I
\p_{\bar{2}}^J) ,
\eeq
where $(T^{(i)})^a_{IJ}$ stands for a generator of the flavour group
$SU(n_i)$ for
the representation $R_i$ and index $a$ corresponds to an adjoint
representation of the $SU(n_i)$ group.
It is easy to check that these currents
are $Q$ closed at the classical level and
non-anomalous at the quantum level.
It is easy also to check that they are also holomorphic in
the $Q$-cohomology.
In the massless theory they together with the non-anomalous singlet ones
generate a Kac-Moody algebra that
corresponds to a flavour group symmetry of the theory.
The defining relations for such a Kac-Moody algebra can be calculated in
the weak coupling limit since the action is $Q$-exact.
Hence the calculation can be done in a model of free chiral
supermultiplet,
while the space of the physical \op s is restricted by the $Q$-cohomology
in the interacting theory.

By using the techniques of section 3 it is easy to check that the linear
combinations of the currents $J_{(i)}^a$ which
correspond to the group given by \eq{fgroup} indeed generate the
Kac-Moody algebra with the $G_F$ structure constants
\beq
J_{(i)}^a (z,\bar{z}) J^b_{(i)} (0,0) = \frac{k_i \delta^{ab}}{z^2} +
\frac{if^{abc}}{z} J^c_{(i)} (0,0) +
\eeq
$$ +{\rm regular \;\; terms} \; + \; Q-{\rm exact \;\; terms} ,$$
where $z$ and $\bar{z}$ are the complex coordinates on $\Sigma_1$ and
the level $k$ is given by
\beq
k_i = -\c \; T(R_i)
\eeq
($\c$ is the Euler characteristic of $\Sigma_2$ and $T(R_i)$ stands for
Dynkin index of representation $R_i$).
Notice that the level of this Kac-Moody algebra is negative for
$\Sigma_2 = S^2 ,$ zero for $\Sigma_2 = T^2$ and
and positive for higher genera.

We now construct a stress tensor on $\Sigma_2$ which is $Q$-closed and
holomorphic in the $Q$-cohomology.
Let us introduce the following gauge invariant \op \
\beq
T(z,\bar{z})= T_g (z,\bar{z}) +T_m (z,\bar{z}),
\label{strtot}
\eeq
where
\beq
T_g (z,\bar{z}) =
\frac{2\pi}{e^2} \sg2 d^2 u[2g^{\bar{2} 2} F_{12} F_{1\bar{2}} - i (D_1
\bl) \c_1 ] ,
\label{strg}
\eeq
\beq
T_m (z,\bar{z}) =2\pi \sg2 d^2 u [-(D_1\bh)(D_1\h) + g^{2\bar{2}} \bp_{12}
(D_1 \p_{\bar{2}})] .
\label{strm}
\eeq
Here $z=z^1$ and $\bar{z} =
z^{\bar{1}}$ are coordinates on the surface $\Sigma_1.$
We imply that the matter fields in \eq{strm} belong to the (reducible)
representation $R,$ i.e. all the present matter fields contribute to
\eq{strm}.
At the classical level this \op \ $T(z,\bar{z})$ obeys
\beq
\{ Q,T\} = 0,\;\;\; \partial_{\bar{1}} T = \{ Q, ...\} .
\label{gholom}
\eeq
When the gauge is fixed by adding $L_{\rm fix}$ (see \eq{Lfix})
we have to modify the stress tensor.
Thus instead of \eq{strg} we introduce $T_g^{qu}$
\beq
T^{qu}_g (z,\bar{z}) =
\frac{2\pi}{e^2} \sg2 d^2 u[2g^{\bar{2} 2} F_{12} F_{1\bar{2}} - i (D_1
\bl) \c_1 -(D_1^{(0)}B) A_1 + (D^{(0)}_1 c^+)(D_1 c)]  ,
\label{strgqu}
\eeq
and
\beq
T(z,\bar{z})= T_g^{qu} (z,\bar{z}) +T_m (z,\bar{z}).
\label{strtqu}
\eeq
By using the equations of motion \rf{eqmotion} and \rf{mattermotion}
one can check that eqs.~\rf{gholom} are still valid if we use the total
BRST charge $Q' =Q +Q_{BRST}$ (see Eq.(5.1)) instead of $Q.$

One may therefore expect that the \op \ algebra generated by $T$ is
holomorphic in the $Q$-cohomology.
Actually we should still check that there is no anomaly at the quantum level
which could spoil the $Q$-closeness and holomorphicity of $T.$
Such an anomaly could appear due to the presence of the gauge
interactions.
By an explicit calculation we shall demonstrate (at the one loop level)
that the ``anomalous'' contributions coming from $T_g$ and $T_m$ cancel
each other.
To this end we have to introduce a gauge invariant ultraviolet
regularization.
For simplicity we shall consider the case of $H^{2,0}\neq 0.$
In this case there exists a non-trivial $(2,0)$ holomorphic form on $M.$
As a consequence it allows us
to introduce the Pauli-Villars ultra-violet regularization.
At the one loop level in the background gauge such a
regularization is sufficient to maintain the gauge invariance of the
theory (for similar anomaly calculations see, e.g. \cite{anj}).

In our one-loop approximation we shall calculate the matrix elements of
the \op s $\{ Q,T\}$ and $\partial_{\bar{1}} T$ in external gauge
($A_{cl}^{\mu}$) and
gluino ($\c_n^{cl}$) fields.

In the quadratic approximation the Lagrangian
for physical quantum fields read
\beq
L =\sqrt{g} \frac{1}{e^2} {\rm Tr} [-2A^n (D^m D_m
\delta_n^k +i F^k_n -iF^p_{\; p}\delta_n^k ) A_k +
iA_m F^{mn} A_n +iA^m F_{mn} A^n +
\label{quadrphys}
\eeq
$$+i \bar{\lambda}^{mn} D_m \chi_n +\bar{\lambda}^{mn} A_m
\chi_n^{cl} + i\bar{\lambda} D^n \chi_n +
\bar{\lambda} A^n \chi_n^{cl} -c^+ (D^mD_m +D_m D^m)c - c^+
D_{\bm}^{(0)} \c^{\bm}] + $$
$$+\sqrt{g} ( \bar{\phi} D^{\bar{m}} D_{\bar{m}} \phi +
\bar{\psi}^{\bar{m}\bar{n}} D_{\bar{m}} \psi_{\bar{n}} + \bar{\psi}
D^{\bar{m}} \psi_{\bar{m}} + N_{\bar{m}\bar{n}}
\bar{N}^{\bar{m}\bar{n}} -i\bh \c_n \p^n -e^2 \bh F_{\; m}^m \h ) .$$
Here $F_{\bm m},$ $F^{mn}$ and $F_{mn}$ are the components of the
strength of an external gauge field $A_{\mu}^{cl} ,$ and the
covariant derivatives are taken in the external gauge field.

The Lagrangian for the regulator fields read
(a subscript $reg$ stands in order to label the regulator fields)
\beq
L_{reg} =L_{reg}^g + L_{reg}^m ,
\eeq
where
\beq
L_{reg}^g = \sqrt{g} \frac{1}{e^2} {\rm Tr} [-2A^n_{reg} (D^m D_m
\delta_n^k +iF^k_n -iF^p_{\; p}\delta_n^k )A_k^{reg} +
iA^{reg}_m F^{mn} A^{reg}_n +iA^m_{reg} F_{mn} A^n_{reg} +
\label{regLg}
\eeq
$$+i \bar{\lambda}^{mn}_{reg} D_m \chi_n^{reg} +
\bar{\lambda}^{mn}_{reg} A_m^{reg}
\chi_n^{cl} +
i\bar{\lambda}_{reg} D^n \chi_n^{reg} +
\bar{\lambda}_{reg} A^n_{reg} \chi_n^{cl} - $$
$$-c^+_{reg} (D^mD_m +D_m D^m)c_{reg}-c^+ D_{\bm}^{(0)} \c^{\bm}] +$$
$$+ \sqrt{g}\frac{1}{e^2} {\rm Tr} \left(
\frac{M}{2} E_{mn} \bl^{mn}_{reg} \bl_{reg} + \frac{M}{2} S_{\bm\bn}
\c_{reg}^{\bm}\c_{reg}^{\bn} + 2M^2 A^n A_n + 2M^2 c_{reg}^+ c_{reg} \right)$$
and $L^{reg}_m$ is given by \eq{regm}.
For simplicity we shall restrict ourselves to the case of the forms
$E_{mn}$ and $S_{\bm\bn}$ being non-zero everywhere (this corresponds to the
case $\Sigma_1=\Sigma_2 =T^2$) and $E_{12} S^{12} = -1.$
These mass terms are gauge invariant under gauge
transformations of the background and regulator fields.
However the mass terms for the regulator fields of the gauge multiplet
are not BRST invariant (the regulator fields are assumed to
transform in the same manner as the physical ones under BRST
transformations).
We shall see that despite of this fact no anomaly appears in the relations
\rf{gholom} for the stress tensor $T.$

The regularized stress tensor $T_{reg}$ is defined as a difference of the
stress tensor of the physical fields and that of the regulator ones as
given in eqs.~\rf{strtot}, \rf{strg}, \rf{strm}.
Due to the presence of regulator fields eqs.\rf{gholom} are not now
correct.
Instead we have
\beq
\{ Q' , T_{reg} \} = -2\pi M \sum_{IJ} a_{IJ} \sg2 d^2 u \; E_{12}
\partial_1 (\p_{reg}^{I,2} \h^J_{reg}) +
\label{regQT}
\eeq
$$+\frac{2\pi}{e^2}\sg2 d^2 u {\rm Tr} [-2iM E_{12} (D_1 A_{reg}^2 - D^2
A_{reg,1}) \bl_{reg} +2M^2 A_{reg,1} \c_{reg,1}] .$$
The contribution from the matter sector (the first term on
the right hand side of \eq{regQT}) can be calculated similarly to
the case of an anomaly in the current $J^{(I)} .$
Integrating over the regulator fields we get for this contribution
the following expression
\beq
-\frac{1}{4\pi} \partial_1 \int_{\Sigma_2}
d^2 u {\rm Tr}_R (\c_2 F_{\bar{2} 1} - \c_1 F_{\bar{2} 2}) ,
\label{matan}
\eeq
where we have omitted the subscript $cl$ for external (low energy) fields.

Let us consider now the contribution from the gauge sector, i.e.
the terms in the second line of \eq{regQT}.
By using appropriate vertices in \eq{regLg} we have for that contribution
\beq
\frac{1}{e^2} <4\pi iM \sg2 d^2 u \; E_{12} (D_1 A_{reg}^2 - D^2
A_{reg,1}) \bl_{reg} (z,\bar{z} ,u,\bar{u})
\label{gaugean}
\eeq
$$ \int d^4 y {\rm Tr}
\bl^{12}_{reg} [A_{reg,1} \c^{cl}_2 -A_{reg,2} \c^{cl}_1](y) -$$
$$-4\pi M^2 \sg2 d^2 u\;
A_{reg, 1} \c^{cl}_1 \int d^4 y \bl [A_{reg,\bar{1}}\c^{cl}_1 +
A_{reg,\bar{2}}\c^{cl}_2](y)> .$$
By using the propagators given in Appendix and by expanding the expression
\eq{gaugean} in powers of the external gauge field and gaugino $\c_n ,$
it is straightforward to calculate the contribution of the gauge sector
into $\{ Q' , T_{reg} \} .$
It turns out to be exactly that which is given by \eq{matan} with an
opposite sign.
Thus the anomalous contributions to $<\{ Q' , T_{reg} \}>$ cancel and the
\op \ $T$ is BRST closed at the quantum level.

In a similar way we can check that this \op \ is holomorphic in
the $Q'$-cohomology at the quantum level due to a cancellation of anomalous
contributions from the matter and gauge sectors.
We have
\beq
\partial_{\bar{1}} T_{reg} =
\eeq
$$=- \frac{2\pi}{e^2} \sg2 d^2 u {\rm Tr}
[-M E^{\bar{1}\bar{2}} \bl_{reg} D_1
\bl^{reg}_{\bar{1}\bar{2}} -iM^2 \bl_{reg} \c_1^{reg} - 2M^2 A^{reg}_n (D^n
A_1^{reg} - D_1 A^n_{reg})] +$$
$$+ 2\pi\partial_1
\sg2 d^2 u [ M^2 \sum_I\bh_{reg}^I \h_{reg}^I -
ME^{\bar{1}\bar{2}} \sum_{IJ} a_{IJ}
\p_{\bar{1}}^{reg,I} \p_{\bar{2}}^{reg,J}] + \{ Q,...\} .$$
We actually can calculate the individual contributions from the sectors of
matter fields and of the gauge multiplet in an easier way by observing
that the \op \ in \eq{matan} is just the \op \ $H^{(1)}_{12,\bar{2}}$
integrated over $\Sigma_2 .$
Hence by using eq.(2.26) we get, e.g. that the individual contribution of
the matter sector is given by $H^{(2)}_{12,\bar{1}\bar{2}} /(2\pi) .$
The contribution of the gauge sector comes with an opposite sign and
cancels the contribution of matter fields.
Therefore the \op \ $T$ is holomorphic in the $Q'$-cohomology at the quantum
level.

\subsection{Virasoro algebra on $\Sigma_2$}

We can now consider an \op \ algebra generated by the \op \ $T.$
Since the action of the theory is BRST exact (we consider the massless
theory) the \op \ algebra generated by $T$ can be determined in the weak
coupling limit.
In this limit the \op \ $T$ is a sum of non-interacting stress tensors
for the matter $T_m$ and for the gauge multiplet $T_g^{qu}$
(we consider the usual $\alpha$ gauge fixing without background
gauge field).
The gauge multiplet is effectively reduced to a superposition of
${\rm dim} \; G$ abelian gauge supermultiplets and we take into account
only a quadratic part of $T_g^{qu}$
\beq
T_g^{qu} = \frac{2\pi}{e^2}
\sg2 d^2 u {\rm Tr}[2(D_1 A_2 -D_2 A_1) (D_1 A_{\bar{2}} -
D_{\bar{2}} A_1) -
\label{quadr}
\eeq
$$-i(D_1 \bl) \c_1 - A_1 D_1 (D^n A_n + D_n A^n )
+D_1 c^+ D_1 c],$$
where
$D_n$ and $D^n$ stand for covariant derivatives in the external
gravitational field; we also used in \eq{quadr} an expression for the
field $B$ in terms of the gauge potential
that results from the equations of motion.

For the \op \ $T_m$ it is easy to extract from section 3 that its
contribution to the OPE is given by
\beq
T_m (z,\bar{z}) T_m (w,\bar{w}) = \frac{\c {\rm dim} R}{2(z-w)^4} + \frac{2
T_m}{(z-w)^2} + \frac{\partial_z T_m}{z-w} + {\rm regular \;\;terms} +
\{ Q,...\},
\eeq
where $z$ and $w$ are complex coordinates on $\Sigma_1$ and $\c$ is the
Euler characteristic of $\Sigma_2 .$

Similarly with the calculation of OPE given in section 3 one can
find the OPE for the \op \ $T_g^{qu} .$
It is easy to check that the contributions to the central term
$T_g^{qu} (z,\bar{z}) T_g^{qu} (w,\bar{w})$ of the two last terms in
\eq{quadr} cancel each other.
Thus for the central term in the OPE we essentially have
\beq
<T_g^{qu} (z,\bar{z}) T_g^{qu} (w,\bar{w})> = 4\pi^2
\sg2 d^2 u \sg2 d^2 v
\eeq
$$\left[ 4<(D_1 A_2 (z,u)) (D_1 A_{\bar{2}}(w,v))>
\cdot <(D_1 A_2 (w,v))(D_1 A_{\bar{2}}(z,u))> +\right.$$
$$\left. + <D_1 \bl (z,u)
\c_1 (w,v)> \cdot <D_1 \bl (w,v)\c_1 (z,u)>\right] ,$$
where $z,w$ and $u,v$ are the complex coordinates on the $\Sigma_1$ and
$\Sigma_2$ surfaces respectively.
By using the propagators given in Appendix (with $M=0$ and without an external
gauge field) and the techniques of section 3 we get for the central term in OPE
\beq
<T_g^{qu} (z,\bar{z}) T_g^{qu} (w,\bar{w})> =\frac{-\c {\rm dim}
G}{2(z-w)^4}.
\eeq
The sign ``-'' at the central term appears here in contrast to the
contribution of the matter sector because the vector field
($A_{\bar{2}}$) on $\Sigma_2$ has an opposite statistics as compared to
the field $\p_{\bar{2}} .$
Similarly one can calculate the less singular terms in the OPE.
Thus we get
\beq
T (z,\bar{z}) T (w,\bar{w}) = \frac{c}{2(z-w)^4} + \frac{2
T}{(z-w)^2} + \frac{\partial_z T}{z-w} + {\rm regular \;\;terms} +
\{ Q,...\},
\eeq
where
\beq
c=\c  ({\rm dim}\; R -{\rm dim}\; G).
\eeq
Thus we see that the \op \ $T$ generates a Virasoro algebra in the
$Q$-cohomology with the central charge $c$ which coincides with that
which we can expect from the expression (2.39) for the gravitational
anomaly.

The currents $J^a_{(i)}$ which generate the Kac-Moody algebra are
the primary fields with respect to this stress tensor and have a
conformal dimension 1
\beq
T (z,\bar{z}) J^a_{(i)} (w,\bar{w}) = \frac{1}{(z-w)^2} J^a_{(i)}
(w,\bar{w}) +\frac{1}{z-w}\partial_1 J^a_{(i)}
(w,\bar{w}) + {\rm regular \;\; terms} + \{ Q, ...\} .
\eeq
We again observe that a different Virasoro algebra on $\Sigma_2$
can be also constructed.
Its central charge is equal to the Euler characteristic of the surface
$\Sigma_1 .$
Hence we have a direct sum of two Virasoro algebras.

\subsection{Instanton effects}

In four dimensional supersymmetric gauge theories without a superpotential
for the matter supermultiplet the vacuum state is degenerate due to a valley
in the potential.
This degeneracy is not lifted in perturbation theory due to the
non-renormalizability theorem \cite{grisaru}.
However instantons can induce an effective superpotential and lift a
degeneracy of the vacuum \cite{affleck,vs2,amati}.

In the heterotic topological gauge theory instantons do not induce any
superpotential just because it is forbidden by the Lorentz invariance of the
theory.
Therefore this theory has a moduli space of ground states.
Actually the absence of an effective superpotential in this case
is a counterpart to the analysis of a possible structure of
an effective superpotential in untwisted supersymmetric theories
\cite{affleck,vs2,amati,seiberg,seib} where a superpotential does not
appear for such a particular representation of the matter fields
which enter the heterotic topological Yang-Mills theory
due the global phase symmetries of the theory.
On a curved manifold the dynamics could differ from that of the theory
in a flat space-time.
However for the Wilsonian effective action that we shall consider in
this section it is natural to assume that the main conclusions drawn for
the untwisted supersymmetric
theory in flat space-time remains true because the
Wilsonian action is determined by ultraviolet contributions
\cite{vashi}.

For simplicity let us consider the supersymmetric massless gauge theory
with the $SU(2)$ gauge group
and four copies of the fundamental representation for the matter fields
(the mixed anomaly in the twisted version of this theory
is cancelled, see Appendix).
This is just the case \cite{affleck} when no
effective superpotential is induced by instantons and hence the
classical vacuum degeneracy is not lifted.
Therefore the quantum theory has a moduli space of ground states
\cite{seiberg}.

An effective low-energy theory can be described in terms of ``mesons''
constructed as gauge invariant composite chiral fields \cite{seiberg}
$V^{ij} = Q^i Q^j ,$ where $Q^i$ stands for a ``quark'' matter chiral
superfield;
indices $i,j =1,2,3,4$ correspond to flavours of ``quarks''.
In the massless theory the flavour group of symmetry is $SU(4).$
At the classical level the Pfaffian of the matrix $V^{ij}$ trivially vanishes
by Bose symmetry.
However due to instantons it acquires a non-zero vacuum expectation value
\beq
{\rm Pf} V = \Lambda^4 ,
\label{Pfvev}
\eeq
where $\Lambda$ is a scale of the gauge theory.
This vacuum expectation value corresponds (in an appropriate basis) to
\beq
V= \Lambda \left( \begin{array}{cc} \sigma_2 & 0 \\ 0 & \sigma_2
\end{array} \right)
\label{vev}
\eeq
and breaks the group symmetry $SU(4) \to Sp(4)$ \cite{seiberg}.
The low energy fields are the fluctuations of $V$ around the expectation
value \eq{vev} subject to constraint \eq{Pfvev}.
Thus the low energy fields belong to $SU(4)/Sp(4)= SU(3)/SU(2) = S^5 .$

We assume that the glueball field $S={\rm Tr} W^2 ,$ where $W$ is the
superstrength of the gauge supermultiplet corresponds to
heavy excitations \cite{seiberg}.
In order to incorporate \eq{Pfvev} into a low energy effective theory one
may use a Lagrange multiplier field $X$ with the superpotential
\cite{seiberg}
\beq
W_{eff} = X({\rm Pf} V - \Lambda^4).
\label{spot}
\eeq
In the twisted theory a superfield $X$ becomes a $(2,0)$-form while
the $F$ term $[W_{eff}]_F$ becomes a $(2,2)$-form.
This $F$ term is $Q$-closed in the twisted theory.
For example the glueball superfield $S$ can play the role of the Lagrange
multiplier field $X.$
In such a case the twisted version of the $F$ term for the effective
potential is given by the $Q$-closed \op \eq{Omix}.
Let us denote
\beq
V^{12}=x_1 , \;\;\; V^{13} =x_2 ,\;\;\; V^{14} =x_3 ,\;\;\;
V^{34} =y_1 ,\;\;\; V_{24} =y_2 ,\;\;\; V^{23} =y_3 ,
\eeq
and put $\lambda =1.$
Then the superpotential \eq{spot} can be rewritten as follows
\beq
W_{eff}= X(\sum_{i=1}^3 x_i y_i -1).
\label{xypot}
\eeq
We now want to determine the ground ring of local scalar \op s
constructed out of the matter chiral fields $x_i$ and $y_i .$
By solving the equations of extremum for the superpotential \eq{xypot}
we get $X=0$ and \eq{Pfvev}.
The ground ring is generated by $x_i$ and $y_i ,$ $i=1,2,3,$
modulo \eq{Pfvev} (this construction is similar to the chiral rings in
two-dimensional $N=2$ supersymmetric theories \cite{LVW}).
Since the Witten index is 2 for the $SU(2)$ supersymmetric gauge theory
\cite{windex}
such a ground ring should be generated by two elements.
Actually the massless limit of the theory that we considered up to now is
not well defined in the following sense.
In the massless theory
we can take any constant matrix conjugate to that of \eq{vev} for a
vacuum expectation value of the field $V^{ij} .$
In turn if we consider a massive theory with small mass parameters
$m_{ij}$ ($m_{ij}< \Lambda$)
\beq
\sum_{i,j=1}^4 m_{ij} V^{ij}
\eeq
then the above degeneracy for the vacuum expectation value of $V^{ij}$
is lifted \cite{amati}.
Hence the massless limit depends on the way the masses vanish.

Therefore in order to control the behaviour of the theory we add
masses to the fields $Q^i$ and hence for the effective potential we get\
\beq
W_{eff}= X(\sum_{i=1}^3 x_i y_i -1) +\sum_{i=1}^3 (a_i x_i +b_i y_i),
\label{potmas}
\eeq
where $a_i$ and $b_i$ are mass parameters.
At the extremum of the potential \rf{potmas} we get
\beq
\sum_i x_i y_i -1 =0,\;\;\; Xx_i +b_i=0,\;\;\; Xy_i +a_i =0.
\label{exmas}
\eeq
In the twisted version of the theory
the expressions given in \eq{exmas} are $Q$-exact
by the equations of motion (we imply that $X,$ $x_i$ and $y_i$ are the
lowest components of chiral supermultiplets).

It is easy to see that from \eq{exmas} follows
\beq
X^2 = \sum_i a_i b_i .
\eeq
Let us take for definiteness $a_1 =b_1 =\beta ,$ $a_2=a_3=b_2=b_3=0.$
Then we have
\beq
x_2=x_3=y_2=y_3 =0,
\;\;\; X=-\beta x_1 =-\beta y_1 ,\;\;\; X^2 =\beta^2 .
\eeq
Thus the ground ring is generated by the \op s $1,\; X$ modulo $X^2 =\beta^2
.$
It is easy to see that if we do not take into account the instanton
effects then we should put $\Lambda =0$ in the superpotential \eq{spot}
and the ground ring is generated by the \op \ $1,\; X$ modulo $X^2 =0.$

The ground ring in this model
is quite similar to the one in $N=2$ supersymmetric
$CP^1$ models \cite{cpvafa}.
Moreover \cite{zee,witten2,phase,cpvafa}
the effect of instantons in the $CP^1$ supersymmetric sigma model
is exactly to change the \cn \ $X^2=0$ into $X^2 =\beta^2 .$
Because in our heterotic topological theory we have embedded conformal
theories in $Q$-cohomology we may conjecture that there is a
correspondence between instantons and renormalization group in four
dimensional supersymmetric theory with this particular choice of
representation for the matter fields and those in an appropriate
two dimensional $N=2$ supersymmetric sigma model \cite{joh1}.
A half-twisted version of such a sigma model is to have a chiral
conformal stress tensor with the same central charge as the
chiral conformal tensor in the twisted 4D theory.
In the present particular example the central charge $c=5\c .$

We may also conjecture that there exists a relation between the moduli space
of heterotic topological theories and that of two dimensional
$N=2$ supersymmetric sigma models.

\section{Conclusions}
\setcounter{equation}{0}

We have shown that in a twisted $N=1$ SUSY model with a
single free chiral supermultiplet on the four manifold $M=\Sigma_1
\times \Sigma_2$ there exist two chiral
infinite dimensional symmetries $W_{1+\infty}$ in the cohomology of the BRST
\op .
The generators of such an algebra are integrals over $\Sigma_2 (\Sigma_1)$
of the bilinear composite \op s.
The central charge of the $W_{1+\infty}$
algebra is the Euler characteristic $\c_2 (\c_1)$ of $\Sigma_2 (\Sigma_1).$

It is worth noticing that the representation of $W_{1+\infty}$ given
here is very close to a representation of this algebra in terms of
2D free (fermionic or bosonic) fields \cite{bakas1,depireux,bakas}.
It is amusing however that in our representation the central charge
has a purely geometric origin.

The theory becomes a dynamical one if we introduce a superpotential.
We have shown in a particular example of a non-trivial quasihomogeneous
superpotential for a single chiral supermultiplet that
the two algebras $W_{1+\infty}$ are reduced to
two chiral Virasoro algebras with central charges proportional to
$\c_2 (\c_1).$
We point out at this point that one can try to extend our construction
to a model with any number of chiral supermultiplets \cite{joh1}.
In such a model it will be interesting to see if
a realization of $W_N$ and $W^p_{\infty}$ \cite{bakas1}
algebras can similarly be obtained directly from a
four dimensional quantum field theory.

In the heterotic topological gauge theory (without a superpotential for
the matter multiplet) there exist two chiral Virasoro algebras with
the central charges proportional to $\c_2 (\c_1)$ and a Kac-Moody algebra
in the BRST cohomology corresponding to
the group of the flavour symmetry of the theory with the level
proportional to $\c_2 (\c_1).$

Notice that the cohomology of the BRST \op \ is larger than
the one we discussed in this paper.
We postpone a detailed analysis of it for a next publication.

We also point out that the supersymmetric theory can be twisted
differently as we discussed in section 2.
For such a mirror model we could get an anti-chiral
conformal algebras in the corresponding BRST cohomology with the same
central charges.

It is important that this (extended) conformal structure is invariant
under the renormalization group due to the BRST symmetry.
Therefore it contains an important information on the dynamics of the
untwisted supersymmetric theory.
We demonstrated a similarity of the ground ring structure of the
heterotic topological theory to that of twisted $N=2$ supersymmetric
sigma models.
This fact hints the existence of a relation between the renormalization
group flow and instantons in four dimensional supersymmetric gauge
theories and those in $N=2$ supersymmetric sigma models.
An interesting question is also if this conformal structure corresponds
to a sort of an infrared fixed point in the untwisted four dimensional
supersymmetric gauge theory.

It would also be of some interest to extend such a construction to the
heterotic topological gauge theories with a superpotential for the
matter multiplet \cite{joh1}.

We thus arrive at an interpretation for
the chiral Virasoro algebras in $Q$-cohomology
as algebras on a surface $\Sigma_1$($\Sigma_2$) with a classical metric.
It is tempting to try to extend such an interpretation allowing the
metric on $\Sigma_1$($\Sigma_2$) to be a quantum one.
In such a case we could get a two dimensional quantum gravity extracted from
the four dimensional induced quantum gravity theory.
An important ingredient here would be that the induced gravity action
(gravitational anomaly) is non-renormalizable at the multiloop level.
We therefore may think that such a realization of the two dimensional
quantum gravity in terms of the four dimensional theory on
$M =\Sigma_1 \times \Sigma_2$ is self-consistent.

It would also
be interesting to extend the above construction of the conformal
algebras to the case of a four dimensional K\"ahler
manifold $M$ which is more complicated than a direct product of two
Riemann surfaces.
One may hope that in such a case it will be also possible to recover two
dimensional integrable structures embedded into four dimensional quantum
field theory.
A different way to reveal integrable structures could be
a deformation of a heterotic topological theory which possesses a
conformal structure through the introduction of
a superpotential for the matter fields \cite{joh1}.

\section{Acknowledgments}

I am grateful to M.Axenides, M.B.Halpern, N.Obers,
J.L.Petersen and G.Savvidy
for interesting discussions of the results of
this work and to M.Axenides for a careful reading of the manuscript.
I also thank the high energy group at NBI where
this work was finished for its hospitality.
The present work was supported in part by a NATO grant GRG 930395.

\appendix
\section{Appendix}
\renewcommand{\theequation}{A.\arabic{equation}}
\setcounter{equation}{0}

\subsection{Solution to the \cn \ of cancellation of the mixed anomaly}

The condition \rf{conMix} of cancellation of the mixed anomaly has the
following
solutions.

(A).  $SU(m) = A_{m-1} .$
Table 1 gives all the dimensions and Dynkin indices of the
representations of $SU(m)$ that can occur in heterotic topological
theories.

\begin{tabular}[t]{lllr}
\multicolumn{4}{c}{Table 1} \\
\multicolumn{4}{c}{The dimensions and Dynkin index of $SU(m)$
representations.} \\
\hline
Representation $R_i$ & Dimension & $T(R_i)$ & Range of $m$\\ \hline
$\;$ & $\;$ & $\;$ & \\
$R_1 =\Box $ & $m$ & 1/2 & $2\leq m$ \\
$\;$ & $\;$ & $\;$ & \\
$R_2=\Box$ \hspace{-.59cm} \raisebox{-1.2ex}{$\Box$}
& $m(m-1)/2$ & $(m-1)/2$ & $4\leq m$ \\
$\;$ & $\;$ & $\;$ & \\
$R_3=\Box \hspace{-.1cm} \Box $ & $m(m+1)/2$ & $(m+2)/2$ & $3\leq m$\\
$\;$ & $\;$ & $\;$ & \\
$R_4=\Box$ \hspace{-.45cm}\raisebox{-1.2ex}{$\Box$}

\hspace{-.45cm}\raisebox{-2.4ex}{$\Box$}
& $m(m-1)(m-2)/6$ & $(m-2)(m-3)/4$ & $6\leq m\leq 8$\\
$\;$ & $\;$ & $\;$ & \\
$R_5=(m-1) \Box \hspace{-.1cm}\Box$

\hspace{-.59cm} \raisebox{-1.2ex}{$\Box$}

\hspace{-.45cm} \raisebox{-2.4ex}{.}

\hspace{-.40cm} \raisebox{-3.6ex}{.}

\hspace{-.50cm} \raisebox{-6.0ex}{$\Box$}
& $m^2 -1$ & $m$ & $2\leq m$ \\
$\;$ & $\;$ & $\;$ & \\ \hline
\end{tabular}

The representation $R_5$ stands for the adjoint representation of
$SU(m).$
Let the repetitions of representations $R_i$ be $n_i.$
The following cases are the solutions of \eq{conMix}.

(A1). $n_5 =1,$ $N-1=n_2=n_3=n_4 =0.$ In this case the theory coincides
with Witten's topological Yang-Mills theory and
${\rm dim} R -{\rm dim} G =0 .$

(A2). $n_1=2m,$ $n_2=n_3=n_4=n_5 =0 .$ In this case
${\rm dim} R -{\rm dim} G = 2m^2 -m^2 +1 = m^2 +1.$

(A3). $n_1=m-2,$ $n_2=0,$ $n_3=1,$ $n_4=n_5 =0,$ $m\geq 3.$
In this case ${\rm dim} R -{\rm dim} G = m(m-3)/2 +1.$

(A4). $n_2=p,$ $n_1=2m-p(m-2),$ $n_3=n_4=n_5 =0,$ $m\geq 4.$
In this case ${\rm dim} R -{\rm dim} G = m^2 +1 - pm(m-3)/2 > 0.$

(A5). $n_2=n_3=1,$ $n_1=n_4=n_5=0,$ $m\geq 4.$
In this case ${\rm dim} R -{\rm dim} G = 1.$

(A6). $n_4=2,$ $n_1=n_2=n_3=n_5=0,$ $m =6.$
In this case ${\rm dim} R -{\rm dim} G = 5.$

(A7). $n_1=(9m-m^2-6)/2,$ $n_4=1,$ $n_2=n_3=n_5=0,$ $6\leq m\leq 8.$
In this case ${\rm dim} R -{\rm dim} G = 21$ at $m=6,$ 33 at $m=7$
and 50 at $m=8.$

(A8). $n_1=2,$ $n_2=n_4=1,$ $n_3=n_5=0,$ $m=6.$
In this case ${\rm dim} R -{\rm dim} G = 12.$

(B)  $SO(2m+1)=B_m .$ The representations of this gauge group with the
Dynkin index less than $C_2 (G)$ are shown in table 2.

\begin{tabular}[t]{lllr}
\multicolumn{4}{c}{Table 2} \\
\multicolumn{4}{c}{The dimensions and Dynkin index of $SO(2m+1)$
representations.} \\
\hline
Representation $R_i$ & Dimension & $T(R_i)$ & Range of $m$\\ \hline
$\;$ & $\;$ & $\;$ & \\
$R_1$ fundamental & $2m+1$ & 1 & $2\leq m$ \\
$\;$ & $\;$ & $\;$ & \\
$R_2$ spinor & $2^m$ & $2^{m-3}$ & $2\leq m\leq 6$ \\
$\;$ & $\;$ & $\;$ & \\
$R_3$ adjoint & $m(2m+1)$ & $2m-1$ & $2\leq m$\\
$\;$ & $\;$ & $\;$ & \\ \hline
\end{tabular}

The cases for the solutions of \eq{conMix} are the following.

(B1). $n_3=1,$ $n_1=n_2=0,$ $m\geq 2.$
This theory coincides with Witten's topological Yang-Mills theory and
${\rm dim} R -{\rm dim} G = 0.$

(B2). $n_1=2m-1,$ $n_2=n_3=0,$ $m\geq 2.$
In this case ${\rm dim} R -{\rm dim} G =2m^2-m-1 >0.$

(B3). $n_2=6,$ $n_1=n_3=0$ for $SO(5).$
In this case ${\rm dim} R -{\rm dim} G =14.$

(B4). $n_2=5,$ $n_1=n_3=0$ for $SO(7).$
In this case ${\rm dim} R -{\rm dim} G =19.$

(B5). $n_2=p,$ $n_1= (2m-1)-2^{m-3} p.$
There are the following cases.

$SO(5),$ $n_1=n_2=2,$ ${\rm dim} R -{\rm dim} G =8.$

\vspace{-.2cm}
$SO(5),$ $n_1=1,$ $n_2=4,$ ${\rm dim} R -{\rm dim} G =11.$

\vspace{-.2cm}
$SO(7),$ $n_1=4,$ $n_2=1,$ ${\rm dim} R -{\rm dim} G =15.$

\vspace{-.2cm}
$SO(7),$ $n_1=3,$ $n_2=2,$ ${\rm dim} R -{\rm dim} G =16.$

\vspace{-.2cm}
$SO(7),$ $n_1=2,$ $n_2=3,$ ${\rm dim} R -{\rm dim} G =17.$

\vspace{-.2cm}
$SO(7),$ $n_1=1,$ $n_2=4,$ ${\rm dim} R -{\rm dim} G =18.$

\vspace{-.2cm}
$SO(9),$ $n_1=5,$ $n_2=1,$ ${\rm dim} R -{\rm dim} G =25.$

\vspace{-.2cm}
$SO(9),$ $n_1=3,$ $n_2=2,$ ${\rm dim} R -{\rm dim} G =23.$

\vspace{-.2cm}
$SO(9),$ $n_1=1,$ $n_2=3,$ ${\rm dim} R -{\rm dim} G =21.$

\vspace{-.2cm}
$SO(11),$ $n_1=5,$ $n_2=1,$ ${\rm dim} R -{\rm dim} G =32.$

\vspace{-.2cm}
$SO(11),$ $n_1=1,$ $n_2=2,$ ${\rm dim} R -{\rm dim} G =20.$

\vspace{-.2cm}
$SO(13),$ $n_1=3,$ $n_2=1,$ ${\rm dim} R -{\rm dim} G =25.$

(C). $Sp(2m)=C_m.$
The representations allowed by \eq{conMix} are given in table 3.

\begin{tabular}[t]{lllr}
\multicolumn{4}{c}{Table 3} \\
\multicolumn{4}{c}{The dimensions and Dynkin index of $Sp(2m)$
representations.} \\
\hline
Representation $R_i$ & Dimension & $T(R_i)$ & Range of $m$\\ \hline
$\;$ & $\;$ & $\;$ & \\
$R_1$ & $2m$ & 1 & $3\leq m$ \\
$\;$ & $\;$ & $\;$ & \\
$R_2$ & $m(2m-1)-1$ & $2m-2$ & $3\leq m$ \\
$\;$ & $\;$ & $\;$ & \\
$R_3$ adjoint & $m(2m+1)$ & $2m+2$ & $3\leq m$\\
$\;$ & $\;$ & $\;$ & \\ \hline
\end{tabular}

The cases for the solutions of \eq{conMix} are the following.

(C1). $n_3=1,$ $n_1=n_2=0.$
This theory coincides with Witten's topological Yang-Mills theory and
${\rm dim} R -{\rm dim} G = 0.$

(C2). $n_1=2m+2,$ $n_2=n_3=0.$
In this case ${\rm dim} R -{\rm dim} G =2m^2+3m.$

(C3). $n_1=4,$ $n_2=1,$ $n_3=0.$
In this case ${\rm dim} R -{\rm dim} G =6m-1.$

(D). $SO(2m) =D_m.$
The representations that have the Dynkin index $T(R_i)$ less than or
equal to $(2m-2)$ are given in table 4. As in the case of $SO(2m+1)$
group only the fundamental, spinor and the adjoint representations are
allowed.
Since $SO(4)$ and $SO(6)$ are isomorphic to $SU(2)\times SU(2)$ and
$SU(4),$ only $m\geq 4$ are relevant.

\begin{tabular}[t]{lllr}
\multicolumn{4}{c}{Table 4} \\
\multicolumn{4}{c}{The dimensions and Dynkin index of $SO(2m)$
representations.} \\
\hline
Representation $R_i$ & Dimension & $T(R_i)$ & Range of $m$\\ \hline
$\;$ & $\;$ & $\;$ & \\
$R_1$ fundamental & $2m$ & 1 & $4\leq m$ \\
$\;$ & $\;$ & $\;$ & \\
$R_2$ spinor & $2^{m-1}$ & $2^{m-4}$ & $5\leq m \leq 7$ \\
$\;$ & $\;$ & $\;$ & \\
$R_3$ adjoint & $m(2m-1)$ & $2m-2$ & $4\leq m$\\
$\;$ & $\;$ & $\;$ & \\ \hline
\end{tabular}

The cases for the solutions of \eq{conMix} are the following.

(D1). $n_3=1,$ $n_1=n_2=0.$
This theory coincides with Witten's topological Yang-Mills theory and
${\rm dim} R -{\rm dim} G = 0.$

(D2). $n_1=2m-2,$ $n_2=n_3=0.$
In this case ${\rm dim} R -{\rm dim} G =2m^2-3m>0.$

(D3). $n_2=4,$ $n_1=n_3=0$ for $SO(10).$
In this case ${\rm dim} R -{\rm dim} G =19.$

(D4). $n_1= (2m-2) -p2^{m-4} ,$ $n_2=p,$ $n_3=0 .$
In this case there are the following solutions to \eq{conMix}.

$SO(10),$ $n_1=6,$ $n_2=1,$ ${\rm dim} R -{\rm dim} G =31.$

\vspace{-.2cm}
$SO(10),$ $n_1=4,$ $n_2=2,$ ${\rm dim} R -{\rm dim} G =27.$

\vspace{-.2cm}
$SO(10),$ $n_1=2,$ $n_2=3,$ ${\rm dim} R -{\rm dim} G =23.$

\vspace{-.2cm}
$SO(12),$ $n_1=6,$ $n_2=1,$ ${\rm dim} R -{\rm dim} G =38.$

\vspace{-.2cm}
$SO(12),$ $n_1=2,$ $n_2=2,$ ${\rm dim} R -{\rm dim} G =22.$

\vspace{-.2cm}
$SO(14),$ $n_1=4,$ $n_2=1,$ ${\rm dim} R -{\rm dim} G =29.$

$E_6.$ The solutions of \eq{conMix} are

(1). $R$ is the adjoint representation (${\bf 78},$ $C_2(E_6)= 12$).
This theory coincides with Witten's topological Yang-Mills theory and
${\rm dim} R -{\rm dim} G = 0.$

(2). $R$ is four copies of the fundamental representation (${\bf 27}$)
with Dynkin index 3.
In this case ${\rm dim} R -{\rm dim} G =30.$

$E_7.$ The solutions of \eq{conMix} are

(1). $R$ is the adjoint representation (${\bf 133},$ $C_2(E_7)= 18$).
This theory coincides with Witten's topological Yang-Mills theory and
${\rm dim} R -{\rm dim} G = 0.$

(2). $R$ consists of three copies of the fundamental representation
(${\bf 56}$) with Dynkin index 6.
In this case ${\rm dim} R -{\rm dim} G =35.$

$E_8.$ The only solution is the adjoint representation $R =Ad.$
This theory coincides with Witten's topological Yang-Mills theory and
${\rm dim} R -{\rm dim} G = 0.$

$F_4.$ The solutions to \eq{conMix} are

(1). $R$ is the adjoint representation (${\bf 52},$ $C_2(F_4)= 9$).
This theory coincides with Witten's topological Yang-Mills theory and
${\rm dim} R -{\rm dim} G = 0.$

(2). $R$ consists of three copies of the fundamental representation
(${\bf 26}$) with Dynkin index 3.
In this case ${\rm dim} R -{\rm dim} G =26.$

$G_2.$ The solutions to \eq{conMix} are

(1). $R$ is the adjoint representation (${\bf 14},$ $C_2(G_2)= 112$).
This theory coincides with Witten's topological Yang-Mills theory and
${\rm dim} R -{\rm dim} G = 0.$

(2). $R$ consists of 4 copies of fundamental representation (${\bf 7}$).
In this case ${\rm dim} R -{\rm dim} G =14.$

\subsection{Propagators}

The propagators of the massive fields in the external gauge fields read
\beq
<\h (x) \bar{\h} (y)> = - \frac{1}{\sqrt{g(y)}}
\frac{1}{M^2 -\Delta_{00}} \delta^4 (x-y) ,
\eeq
\beq
<\p_{\bn} (x) \bp_{12} (y)> =
\eeq
$$=\frac{1}{\sqrt{g(y)}}
\left[ \left(\frac{1}{M^2 -\Delta_{0,1}}
\right)_{\bn 1} D_2 - \left(\frac{1}{M^2 -\Delta_{0,1}}
\right)_{\bn 2} D_1 \right] \delta^4 (x-y) ,$$
\beq
<\bp(x)\bp_{12}(y)>= \frac{1}{\sqrt{g(y)}}
ME_{12}\frac{1}{M^2 -\Delta_{00}}\delta^4 (x-y),
\eeq
\beq
<\p_{\bn} (x) \p_{\bm} (y)> =
\eeq
$$=\frac{1}{\sqrt{g(y)}}\left[ \frac{M}{2} \left(\frac{1}{M^2 -\Delta_{0,1}}
\right)_{\bn p} S_{\; \bar{m}}^p
+\frac{M}{2} S_{\bn}^{\: k} \left(\frac{1}{M^2 -\Delta_{1,0}}
\right)_{k\bar{m}} \right]\delta^4 (x-y),$$
\beq
<\bp (x) \p_{\bn} (y)> = -
\frac{1}{\sqrt{g(y)}}\frac{1}{M^2 -\Delta_{00}} D_{\bn} \delta^4 (x-y),
\eeq
\beq
<N_{\bar{1}\bar{2}} (x) \bar{N}^{12} (y)> =\frac{1}{2} \delta^4 (x-y),
\eeq
where
\beq
\Delta_{00}= D^m D_m ,
\;\;\; (\Delta_{01})^{\bn}_{\bar{k}} =D_m D^m \delta^{\bn}_{\bar{k}}
- [D^{\bn} , D_{\bar{k}}] ,
\eeq
$D_n ,$ $D^n$ are the covariant derivatives in the external
gauge and gravitational fields, and
$(1/\Delta_{01})_{\bar{n} n}$ stand for the $(\bn n)$
components of an \op \ $1/\Delta_{01}$ respectively.

The propagators of free massless fields read as follows
\beq
<\h (x) \bar{\h} (y)> =\frac{1}{\sqrt{g(y)}} \frac{1}{\Delta_{00}}
\delta^{4} (x-y) ,
\label{freeprop}
\eeq
$$<\p_{\bar{1}} (x) \bp_{12} (y)> = - \frac{1}{\sqrt{g(y)}}
\left( \frac{1}{\Delta_{01}}\right)_{\bar{1} 1} D_2 \delta^{4} (x-y) ,$$
$$<\p_{\bar{2}} (x) \bp_{12} (y)> = \frac{1}{\sqrt{g(y)}}
\left( \frac{1}{\Delta_{01}}\right)_{\bar{2} 2} D_1 \delta^{4} (x-y) ,$$
where $\Delta_{00}$ and $\Delta_{01}$ stand for the Laplace \op s
acting on $(0,0)$ and $(0,1)$ forms on $M$ respectively.

Let us consider the gauge supermultiplet.
For the gauge field we have
\beq
<A_{\mu} (x)A_{\nu}(y)>=\frac{e^2}{2\sqrt{g(y)}}
\left(\frac{1}{M^2 - \tilde{\Delta}_{0,1}}\right)_{\mu\nu} \delta^4 (x-y) ,
\eeq
where $\tilde{\Delta}_{0,1}$ stands for the \op \ in the quadratic form
in \eq{quadrphys} for the gauge field.
In a particular case of the external anti-instanton field for which
$F_{mn} =F^{mn} = F^p_{\; p} =0$ we have
\beq
<A_{\bn} (x)A_n(y)>=\frac{e^2}{2\sqrt{g(y)}}
\left(\frac{1}{M^2 - \Delta_{0,1}}\right)_{\bn n}
\delta^4 (x-y)
\eeq
while $<A_{\bm} (x)A_{\bn} (y)> =<A_m (x)A_n (y)> =0.$

For the fermionic fields we have
\beq
<\c_n (x)\bl (y)> =\frac{e^2}{\sqrt{g(y)}}
iD_n\frac{1}{M^2 -\Delta_{00}} \delta^4 (x-y),
\eeq
\beq
<\c_n (x)\bl_{\bar{1}\bar{2}} (y)> =
\eeq
$$=\frac{e^2}{\sqrt{g(y)}}
i\left[\left(\frac{1}{M^2
-\Delta_{1,0}}\right)_{n \bar{2}} D_{\bar{1}} -
\left(\frac{1}{M^2 -\Delta_{1,0}}\right)_{n \bar{1}} D_{\bar{2}}
\right]\delta^4 (x-y),$$
\beq
<\c_n (x) \c_m (y)> =
\eeq
$$=\frac{e^2}{\sqrt{g(y)}}
\left[ \frac{M}{2}\left(\frac{1}{M^2
-\Delta_{1,0}}\right)_{n \bar{p}} E_k^{\; \bar{p}} -
\frac{M}{2} E_n^{\; \bar{p}} \left(\frac{1}{M^2
-\Delta_{0,1}}\right)_{\bar{p} n} \right] \delta^4 (x-y),$$
\beq
<\bl (x) \bl_{\bar{1}\bar{2}} (y)> = \frac{1}{\sqrt{g(y)}}
MS_{\bar{1}\bar{2}} (x) \frac{1}{M^2
-\Delta_{00}} \delta^4 (x-y),
\eeq
\beq
<c(x)c^+ (y)>=\frac{e^2}{\sqrt{g(y)}}
\frac{1}{2M^2 -D^mD_m -D_m D^m } \delta^4 (x-y).
\eeq

\end{document}